\documentclass[twocolumn]{aastex63}

\usepackage{CJKutf8}
\usepackage[scaled]{helvet} 
\usepackage{hyperref} 

\usepackage{mathtools}
\usepackage{balance}

\shorttitle{Bifurcation of planetary building blocks}
\shortauthors{Lichtenberg et al.}
\graphicspath{{./}{figures/}}

\versioninfo{\footnotesize Submitted 2 March 2020; Accepted 10 December 2020; Published 21 January 2021 (\href{https://science.sciencemag.org/content/371/6527/365}{Science 371, 6527})}

\usepackage{placeins} 

\begin{document}

\title{\Large Bifurcation of planetary building blocks during Solar System formation}

\correspondingauthor{Tim Lichtenberg}
\email{tim.lichtenberg@physics.ox.ac.uk}

\author[0000-0002-3286-7683]{Tim Lichtenberg}
\affiliation{Atmospheric, Oceanic and Planetary Physics, Department of Physics, University of Oxford, United Kingdom}

\author[0000-0002-9128-0305]{Joanna {Dr{\c{a}}{\.z}kowska}}
\affiliation{University Observatory, Faculty of Physics, Ludwig-Maximilians-Universit\"at M\"unchen, Germany}

\author[0000-0003-4304-214X]{Maria {Sch{\"o}nb{\"a}chler}}
\affiliation{Institute for Geochemistry and Petrology, Department of Earth Sciences, ETH Zurich, Switzerland}

\author[0000-0002-9501-8347]{Gregor J. Golabek}
\affiliation{Bayerisches Geoinstitut, University of Bayreuth, Germany}

\author[0000-0001-6203-4482]{Thomas O. Hands}
\affiliation{Institute for Computational Science, University of Zurich, Switzerland}


\begin{abstract} 
Geochemical and astronomical evidence demonstrate that planet formation occurred in two spatially and temporally separated reservoirs. The origin of this dichotomy is unknown. We use numerical models to investigate how the evolution of the solar protoplanetary disk influenced the timing of protoplanet formation and their internal evolution. Migration of the water snow line can generate two distinct bursts of planetesimal formation that sample different source regions. These reservoirs evolve in divergent geophysical modes and develop distinct volatile contents, consistent with constraints from accretion chronology, thermo-chemistry, and the mass divergence of inner and outer Solar System. Our simulations suggest that the compositional fractionation and isotopic dichotomy of the Solar System was initiated by the interplay between disk dynamics, heterogeneous accretion, and internal evolution of forming protoplanets.
\vspace{1.0cm}
\end{abstract} 


\section*{} 
\label{sec:main}

\vspace{-0.6cm}

Planetary systems, including the Solar System, form by accretion from a protoplanetary disk of gas and dust. Astronomical observations of these disks provide evidence for rapid dust coagulation \citep{2020Natur.586..228S}, show ringed substructure \citep{2018ApJ...869L..41A}, and indicate a decrease in total dust mass with disk age \citep{2016ApJ...828...46A}, to below the total masses in fully-assembled exoplanetary systems \citep{2014MNRAS.445.3315N}. This suggests that planet formation starts early. At the onset of planetary accretion, small dust grains coagulate to form larger aggregates (pebbles). These pebbles drift inward under aerodynamic drag and gravitationally collapse when they reach sufficient local over-density \citep{2015SciA....115109J}, which preferentially produces birth planetesimals of $\approx$100 km \citep{2017Sci...357.1026D}. These bodies provide the seeds for the accretion process, but their direct growth by pebble accretion only becomes efficient after $\gtrsim 10^5$--$10^6$ yr, once they have grown to larger embryos through mutual collisions \citep{2019A&A...624A.114L}. 

Meteorites record planet formation in our own Solar System and constrain the astronomical timescales: radiometric dating of meteorites suggests iron core formation (differentiation) in planetesimals by $\approx$1 Myr after the formation of Ca,Al-rich inclusions (CAIs) \citep{2014Sci...344.1150K}, the oldest known solids that formed together with the proto-Sun \citep{2012Sci...338..651C}. Core formation in first-generation, birth planetesimals is driven by internal radiogenic heating from $^{26}$Al (half-life of $\sim 7 \times 10^5$ yr). Therefore, the time interval between planetesimal formation and differentiation ($\gtrsim 10^5$--$10^6$ yr) increases with later formation time, pointing to the formation of the earliest planetesimals in the inner Solar System $\lesssim$ 0.3 Myr after CAI formation \citep{2014Sci...344.1150K}. Combined, evidence from both astronomy and geochemistry suggests the onset of planet formation during the earliest phases of the solar protoplanetary disk. 

Complementing astronomical evidence for disk substructure, meteorite data indicate spatial heterogeneity in the isotopic composition of planetary materials (\emph{Supplementary Materials}): studies on Ti, Cr, Mo, and other isotope systems show variabilities across individual planetary bodies from different orbits \citep{2008E&PSL.266..233L,2009Sci...324..374T,2011E&PSL.311...93W}. Combined with temporal constraints on the accretion process, these divide the outer and inner Solar System reservoirs into chronologically and spatially distinct populations, also known as the carbonaceous chondrite (CC) and non-carbonaceous (NC) reservoirs, after their representative meteorite classes. The isotopic signatures record the heterogeneous distribution of presolar dust from multiple sites of stellar nucleosynthesis, and trace transport and mixing processes during planet formation. In contrast, the elemental abundances and internal structure of planetary bodies were modified through geophysical evolution during planetary formation \citep{2019NatAs...3..307L}. Earth's nucleosynthetic isotope composition is NC-like and disparate from outer Solar System materials (\emph{Supplementary Materials}), but its depletion pattern of moderately volatile elements is close to CCs \citep{2019NatGe..12..564B}. This has been attributed to impact delivery \citep{2010Sci...328..884S} or addition of pebbles \citep{2018Natur.555..507S}. The inner Solar System planets are depleted in highly volatile elements, such as hydrogen, carbon, and nitrogen, which influenced the availability of surface water and the composition of their atmospheres. This has been linked with the accretion of the terrestrial planets inside the snow line, where water ice is not stable during the disk phase,  and later delivery of volatile-rich, outer Solar System materials \citep{PeslierISSI2018,Raymond2020PlanetaryAstrobiology}.

The observed CC/NC dichotomy has been suggested to result from an early formation of proto-Jupiter $\lesssim$ 1 Myr after CAI formation, which would inhibit the aerodynamic drift of dust grains and prevent initially heterogeneous disk materials from mixing \citep{2017PNAS..114.6712K,2018ApJS..238...11D}. This scenario constrains the chronology and physical mechanism of Jupiter's growth \citep{Alibert18NatAstron}, but potentially contradicts astronomical and geochemical evidence on the accretion process (\emph{Supplementary Materials}).

\section*{Formation of two distinct planetesimal populations} 
\label{sec:disk_main}

\begin{figure}[tb]
   \centering
   \includegraphics[width=0.49\textwidth]{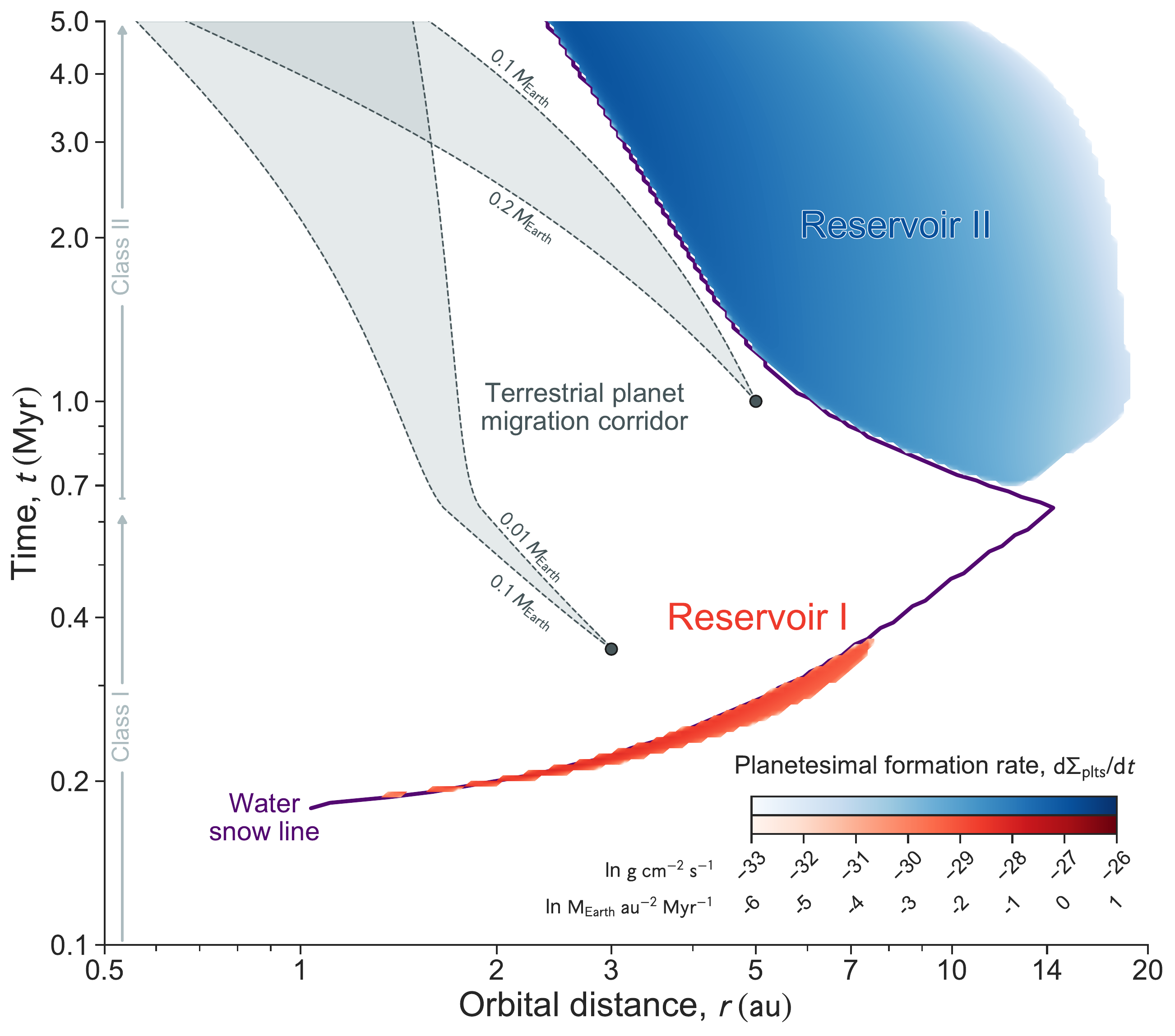}
   \caption{\textsf{\textbf{Formation of two distinct planetesimal populations in the disk simulation.} The rate of planetesimal formation (change in surface density per timestep) is shown on logarithmic scale in cgs and natural units. The solid purple line indicates the orbital migration of the snow line, red and blue areas highlight the formation regions of the two planetesimal populations (Reservoir I and Reservoir II). The dashed gray lines (labeled 'Terrestrial planet migration corridor', cf. Fig.~\ref{fig:s9}) bound the range of possible planet migration scenarios consistent with the present-day orbits of the terrestrial planets. While smaller embryos move slowly inward, more rapidly accreting (and thus more massive) embryos would migrate faster toward the proto-Sun.  The light gray arrows on the left side indicate the transition from the Class I to Class II disk stage.}}
   \label{fig:1}
\end{figure}

Multiple planetesimal populations can be formed during the build-up and evolution of the solar protoplanetary disk \citep{DD18}. We test whether such multi-stage planetesimal formation is consistent with the accretion chronology as inferred from meteorites. We use the output of previous simulations \citep[][see \emph{Supplementary Materials}]{DD18} to explore the CC/NC dichotomy. The simulations cover the initial infall and build-up of gas and dust, during which the disk itself accretes from the surrounding molecular cloud (Class I), and include the drift, coagulation and compositional evolution of the dust, and viscous evolution of the gas during the later evolutionary stage of the disk (Class II). In these simulations, the initial material accretes onto the innermost regions and continuously feeds more distant orbits through viscous expansion of the disk. The snow line moves outward during the Class I stage because a dense, compact disk forms, which heats up viscously. With the onset of the Class II stage, the gas density decreases, the temperature structure becomes dominated by stellar irradiation, and the snow line moves inward again. 

The snow line location affects the redistribution of dust grains. These are initially well-mixed with the disk gas, but radially drift  due to coagulation and aerodynamic drag from the ambient gas. Inward-drifting, icy pebbles undergo rapid dehydration and size reduction at the snow line, which reduces their drift velocity and causes a pile-up of solid material. Outward diffusion of water vapor additionally leads to its recondensation onto icy grains beyond the snow line (the cold-finger effect), locally increasing the density of solids. When the conditions for the formation of dense, gravitationally unstable dust filaments (the streaming instability) are met in the simulation, planetesimals are formed. Planetesimal formation in the simulation preferentially occurs around the snow line. At 5 Myr into the simulation, which is the approximate upper lifetime of the solar nebula (\emph{Supplementary Materials}), we assume that the gas disk dissipates and planetesimal formation halts. 

In wind-driven disks with low levels of midplane turbulence the global angular momentum transport is dominated by near-surface layers (\emph{Supplementary Materials}). In this scenario, the outward-inward tacking snow line generates two distinct episodes of planetesimal formation in different orbital regions and time intervals (Fig.~\ref{fig:1}). The first, early-formed planetesimal population (hereafter Reservoir I) is triggered by the cold-finger effect between $\approx$0.2--0.35 Myr at orbital locations between $\approx$1.3--7.5 astronomical units (au) from the inside-out. The second planetesimal population (Reservoir II) mainly arises from large-scale inward-drift and pebble pile-up at the snow line in the Class II stage. This second population starts forming from $\approx$0.7 Myr between $\approx$17--3 au and proceeds from the outside-in. Due to differences in mechanism efficiency and local dust available, the two reservoirs differ substantially in their total formed mass: Reservoir I forms on the order of an Earth mass, Reservoir II forms on the order of a Jupiter mass in planetesimals \citep[][see \emph{Supplementary Materials}]{DD18}. The build-up of both reservoirs depends on the local evolution of the pebble flux, but in turn influences the coagulation and drift of pebbles toward the inner disk (Figs. \ref{fig:s1}--\ref{fig:s3}). We therefore assess the feedback between reservoir formation and continuing planetary accretion.

\section*{Heterogeneous accretion and reservoir separation} 
\label{sec:accretion_main}

\begin{figure}[tb]
   \centering
   \includegraphics[width=0.49\textwidth]{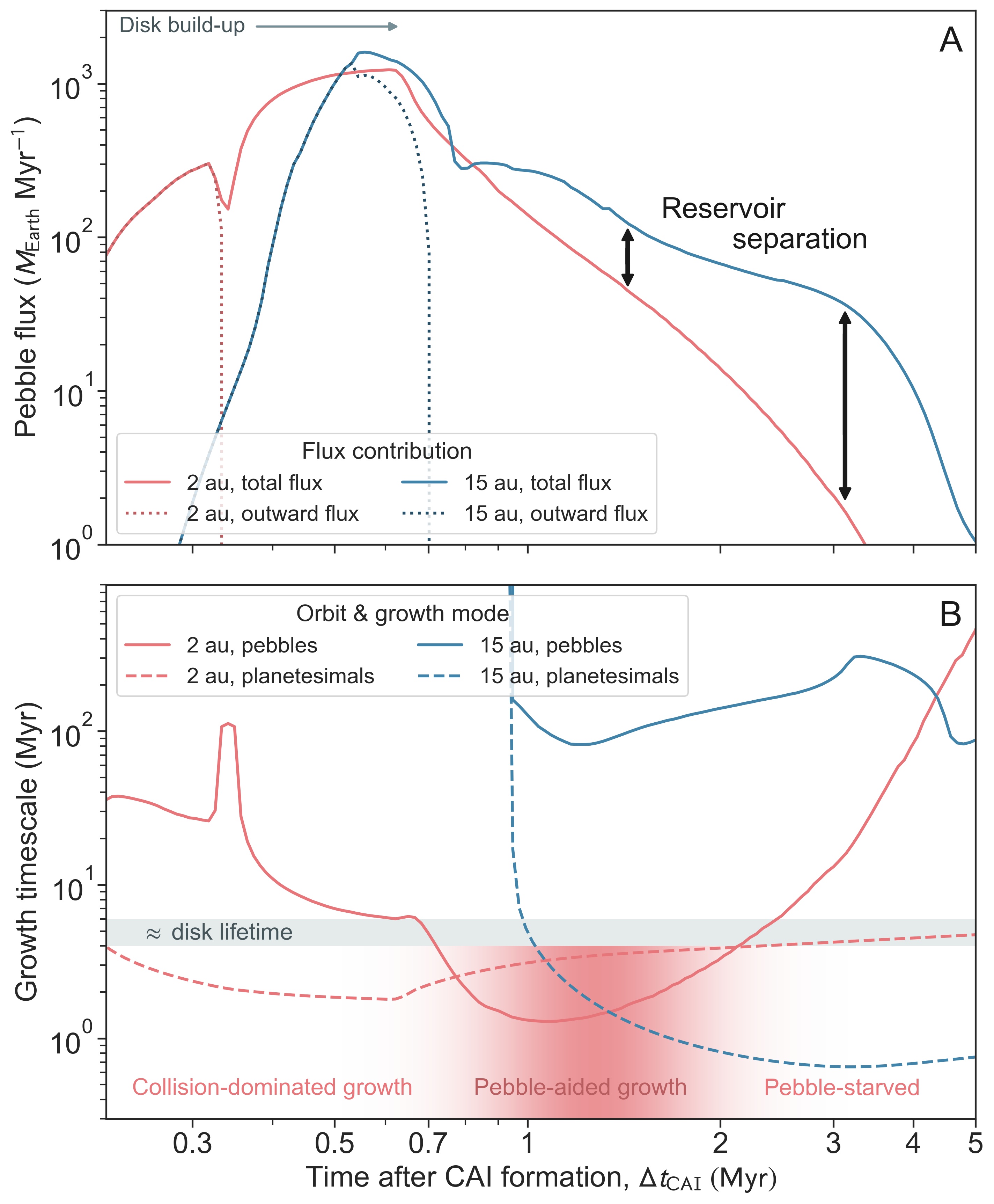}
   \caption{\textsf{\textbf{Pebble flux and planetesimal growth timescales in the disk simulation.} (\textbf{A}) Pebble flux at 2 (red) and 15 (blue) au over time. During the disk build-up stage, the pebble flux is dominated by outward-moving dust (dotted lines), after which the pebbles start drifting inward. Reservoir I and II in the simulation progressively diverge in pebble flux by more than one order of magnitude  due to accretion of Reservoir II (black arrows, labeled 'Reservoir separation'). (\textbf{B}) Comparison of growth timescales for birth-sized planetesimals of 300 km radius by either pebble or planetesimal (collisional) accretion (\emph{Supplementary Materials}) within the approximate lifetime of the solar protoplanetary disk (gray horizontal band, labeled '$\approx$ disk lifetime'). The red shaded area in the bottom indicates the time interval during which pebble accretion is more effective than collisional growth (labeled 'Pebble-aided growth', cf. Fig.~\ref{fig:s11}).}}
   \label{fig:2}
\end{figure}

We aim to determine the dominant mode of solid mass transfer in this disk build-up scenario, and to evaluate its consistency with the timescales of planetary accretion derived from radiometric ages \citep{2010NatGe...3..439R,2011Natur.473..489D,2014Sci...344.1150K}. To relate the numerical simulation to geochemical chronology, we equate time zero in the disk model with the time of CAI formation \citep[][\emph{Supplementary Materials}]{2012Sci...338..651C}. For fixed orbits at 2 and 15 au, approximately representative of Reservoir I and II in the inner and outer disk, we evaluate the anticipated growth timescales for two scenarios of protoplanet accretion (\emph{Supplementary Materials}). We first consider collisional growth due to planetesimal-planetesimal interactions, during which the largest planetesimals grow by accreting smaller bodies. Second, we investigate the efficacy of pebble accretion, which is driven by the abundance of dust grains that cross the protoplanet orbit. Consistent with dynamical evidence from the asteroid-belt \citep{2017Sci...357.1026D} and numerical simulations of the streaming instability \citep{2015SciA....115109J}, we assume a maximum planetesimal radius of 300 km, and that collisional growth is dominated by planetesimals of 50 km in radius. The orbit- and time-dependent evolution of the pebble flux (i.e., the transfer of solid mass in the disk), pebble size, and the gas disk characteristics are self-consistently derived from the disk and coagulation simulation described above \citep[][\emph{Supplementary Materials}]{DD18}.

\begin{figure*}[bth]
  \centering
  \includegraphics[width=0.55\textwidth]{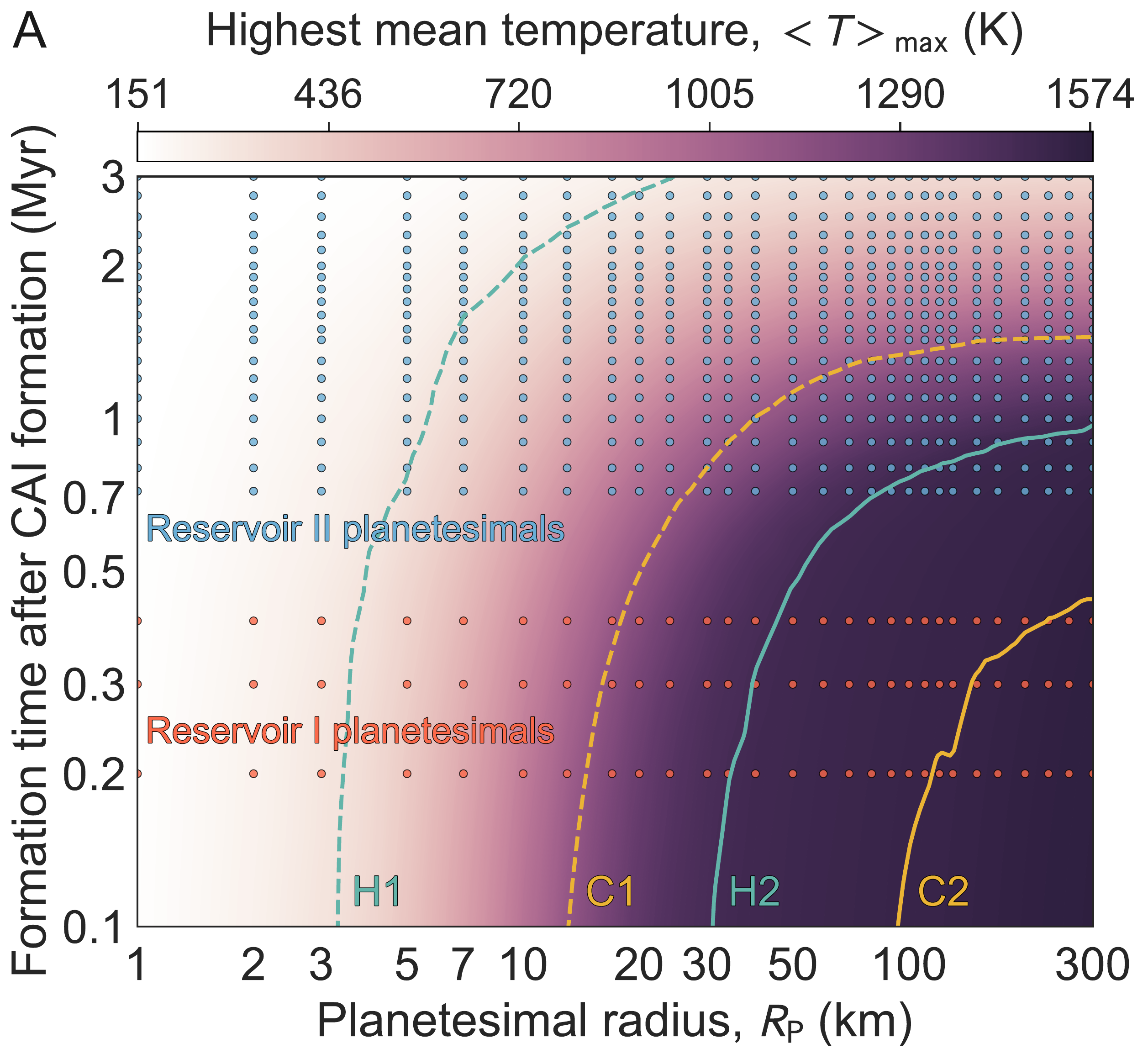}
  \includegraphics[width=0.41\textwidth]{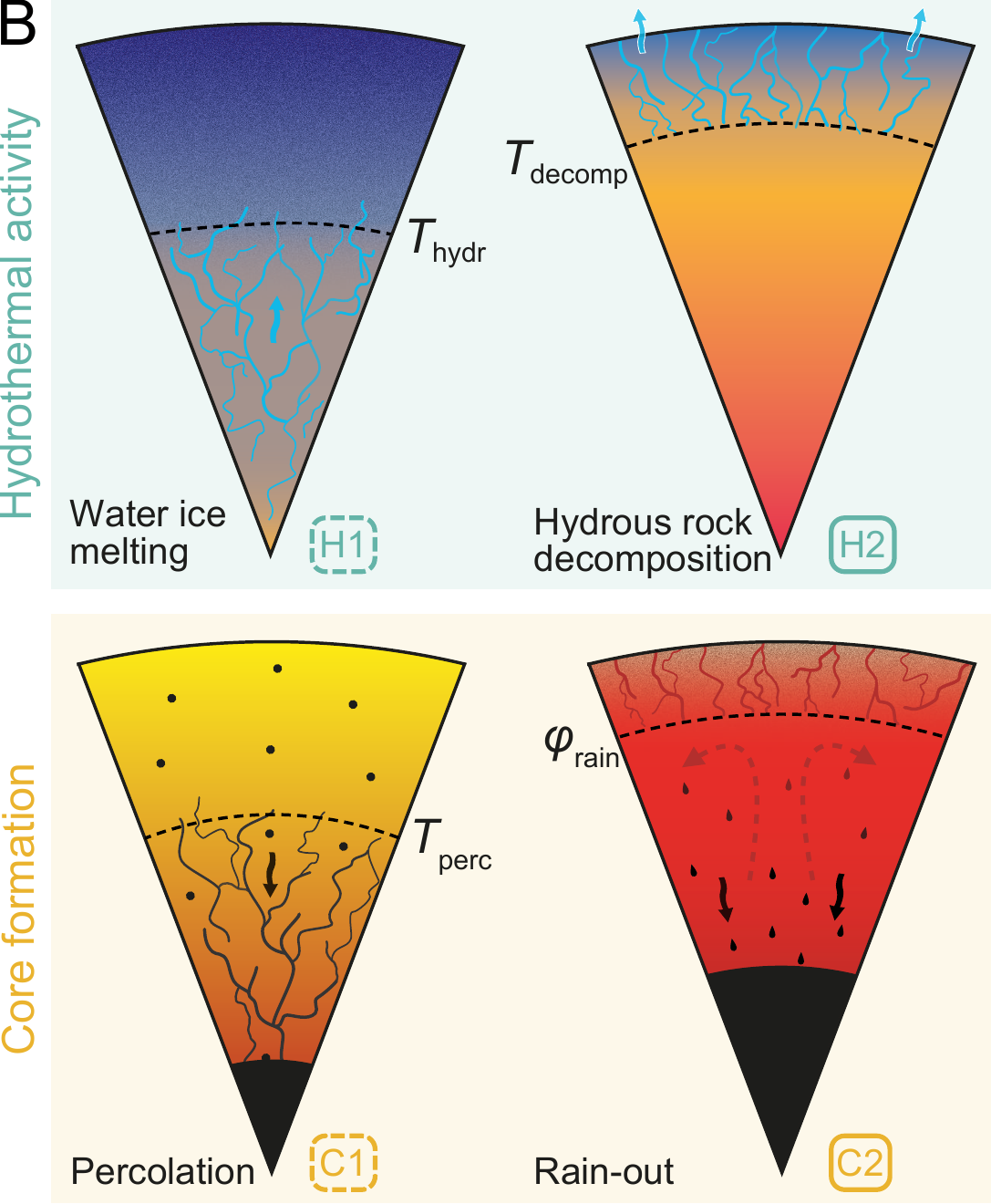}
  \caption{\textsf{\textbf{Simulated thermochemical evolution of planetesimals.} (\textbf{A}) Highest mean internal temperature (color bar), interpolated from simulated planetesimal radii and formation times (dots). Reservoir I and II in the disk model are indicated by red and blue dots, respectively (cf. Fig. \ref{fig:1}). Green and yellow lines (labeled H1--C2) indicate bodies for which 50~vol\% (dashed lines) or 90~vol\% (solid lines) of their interiors undergo hydrothermal activity (H1, H2) and iron core formation (C1, C2). (\textbf{B}) Schematic illustration of the scenarios indicated by the lines in panel \textbf{A} at peak heating and the threshold parameters (see main text and \emph{Supplementary Materials}) to distinguish between regimes. H1, H2: Dark and light blue represent water ice and liquid water, respectively. Silicates are colored yellow to orange, indicating increasing temperature. Dashed black lines indicate the thresholds for water ice melting ($T_\mathrm{hydr}$) and hydrous rock decomposition ($T_\mathrm{decomp}$). The light blue arrows indicate hydrothermal circulation (H1) or degassing of volatiles (H2). C1, C2: Dark gray represents Fe,Ni--S metals. Silicates are colored yellow to red, indicating increasing temperature (yellow to orange) and progressive melting (red). Dashed black lines indicate the thresholds for metal percolation ($T_\mathrm{perc}$) and rain-out ($\varphi_\mathrm{rain}$). Dark gray arrows indicate core formation by percolation (C1) or rain-out (C2). Dark red dashed arrows (C2) indicate convection in the internal magma ocean. }}
  \label{fig:3}
\end{figure*}

Under these conditions, the pebble flux is dominated by outward-moving dust grains (Fig.~\ref{fig:2}A) between $\approx$0.2--0.35 Myr after CAI formation because the disk accretes and viscously expands during the infall stage. The associated outward-directed pebble flux that passes Reservoir I is not efficient enough to drive substantial growth via pebble accretion, but planetesimal-planetesimal interactions lead to substantial growth within the timescale of disk evolution (Fig.~\ref{fig:2}B). Between $\approx$0.35--1 Myr, inward-drifting pebbles reach the inner disk, producing approximately equal contributions to accretion from both pebbles and collisions once the disk transitions into the Class II stage (Fig.~\ref{fig:1}). Subsequently, pebbles substantially contribute to planetesimal growth for a time interval of $\approx$1 Myr.

A final disk phase occurs from $\approx$2 Myr after CAI formation until gas disk dispersal, during which planetesimal growth by pebbles in Reservoir I stalls (\emph{Supplementary Materials}). This is because a substantial fraction of solids, which drift from the outer toward the inner disk, are trapped in Reservoir II, where the pebbles gravitationally collapse to form planetesimals starting from $\approx$1 Myr after CAI formation onward (Fig.~\ref{fig:2}A). The build-up and onset of planetary accretion in Reservoir II shields Reservoir I from drifting grains that originate at larger orbital distances. The difference in pebble flux between the inner and outer Solar System reaches more than an order of magnitude. The pebble flux in the inner Solar System decreases to less than one Earth mass per million years after 3 Myr after CAI formation (Fig.~\ref{fig:2}A). 
\newline
\section*{Divergent geophysical evolution of planetesimal populations} 
\label{sec:planetesimals_main}

These simulation results suggest that both planetesimals and pebbles contribute to planetary accretion. Rocky protoplanets that form from the planetesimal populations in both reservoirs inherit geochemical and geophysical properties from their precursor bodies \citep{2019NatAs...3..307L}. The two planetesimal reservoirs differ substantially in accretion time and thus in radiogenic, internal heating from the decay of short-lived $^{\mathrm{26}}$Al. Meteorite data have shown that $^{\mathrm{26}}$Al heating in early-formed Solar System planetesimals drove their internal evolution, and thus altered their structure and bulk composition \citep[][\emph{Supplementary Materials}]{2017GeCoA.211..115S,2018SSRv..214...36A}. Therefore, we determine the internal and compositional evolution of the Reservoir I and II planetesimal populations formed in the disk simulation. We constrain the timing of iron core formation and hydrothermal activity (aqueous alteration and degassing) using geodynamic simulations of the thermochemical evolution of planetesimals. Our numerical setup assumes isolated planetesimals that form instantaneously and evolve according to a fluid mechanical model \citep[][\emph{Supplementary Materials}]{2019NatAs...3..307L,2018Icar..302...27L}.

We perform 700 single-planetesimal simulations, spanning the parameter range of planetesimal formation time, $t_\mathrm{form} \in$~[0.1, 3.0]~Myr, and planetesimal radius, $R_\mathrm{P} \in$ [1, 300]~km (Fig.~\ref{fig:3}A). To quantify their compositional evolution, we employ four thermochemical criteria (Fig.~\ref{fig:3}B) to determine the onset and cessation of core formation and hydrothermal activity (\emph{Supplementary Materials}). Core formation in planetesimals initiates with the percolation of Fe,Ni--S liquids before silicate melting occurs, and terminates once the interior is sufficiently melted to form an internal magma ocean. At this stage metal droplets rain out from the convecting magma flow, producing complete differentiation between the metal core and liquid silicate mantle \citep{2018Icar..302...27L}. Quantitatively, we assume that core formation occurs in regions with temperatures $T$ higher than the threshold for metal-sulfide percolation, $T_\mathrm{perc} \equiv 1273$ K, and completes for silicate melt fractions $\varphi$ above the rheological transition, $\varphi_\mathrm{rain} \equiv 0.4$, where rocks start to behave like a liquid rather than a solid. As temperature limits on hydrothermal (water-rock) activity, we assume water ice melts above $T_\mathrm{hydr} \equiv 273$ K, and hydrous rock phases fully decompose above $T_\mathrm{decomp} \equiv 1223$ K. 

Averaged over the planetesimal volume, Fig.~\ref{fig:3}A indicates that larger and earlier-formed planetesimals experience higher temperatures during their evolution, leading to a greater degree of internal processing. Compared to Reservoir II, Reservoir I planetesimals reach systematically higher temperatures and undergo body-wide metal-silicate segregation. Reservoir I bodies larger than about 30 km in radius also reach near-complete dehydration. Reservoir II planetesimals only experience limited degrees of heating and therefore less core formation and dehydration. The peak temperatures (Fig.~\ref{fig:3}) only reveal approximate trends in the simulations, as heating and compositional evolution are highly variable on timescales of $10^5$--$10^6$ yr (Figs.~\ref{fig:s4}--\ref{fig:s7}).

To compare the simulations with the meteorite record, we explore the time-dependent evolution of the planetesimal populations in both reservoirs. We assume the planetesimal number distributions follow that of the streaming instability (\emph{Supplementary Materials}) in the interval $R_\mathrm{P} \in$ [1, 300]~km, and take into account the generation of new planetesimals in the simulation. We then evolve the birth planetesimal populations in Reservoir I and II and derive the fraction of each reservoir that falls within the threshold criteria, normalized to the final reservoir population after 5 Myr in the disk simulation.

\begin{figure}[t]
   \centering
   \includegraphics[width=0.49\textwidth]{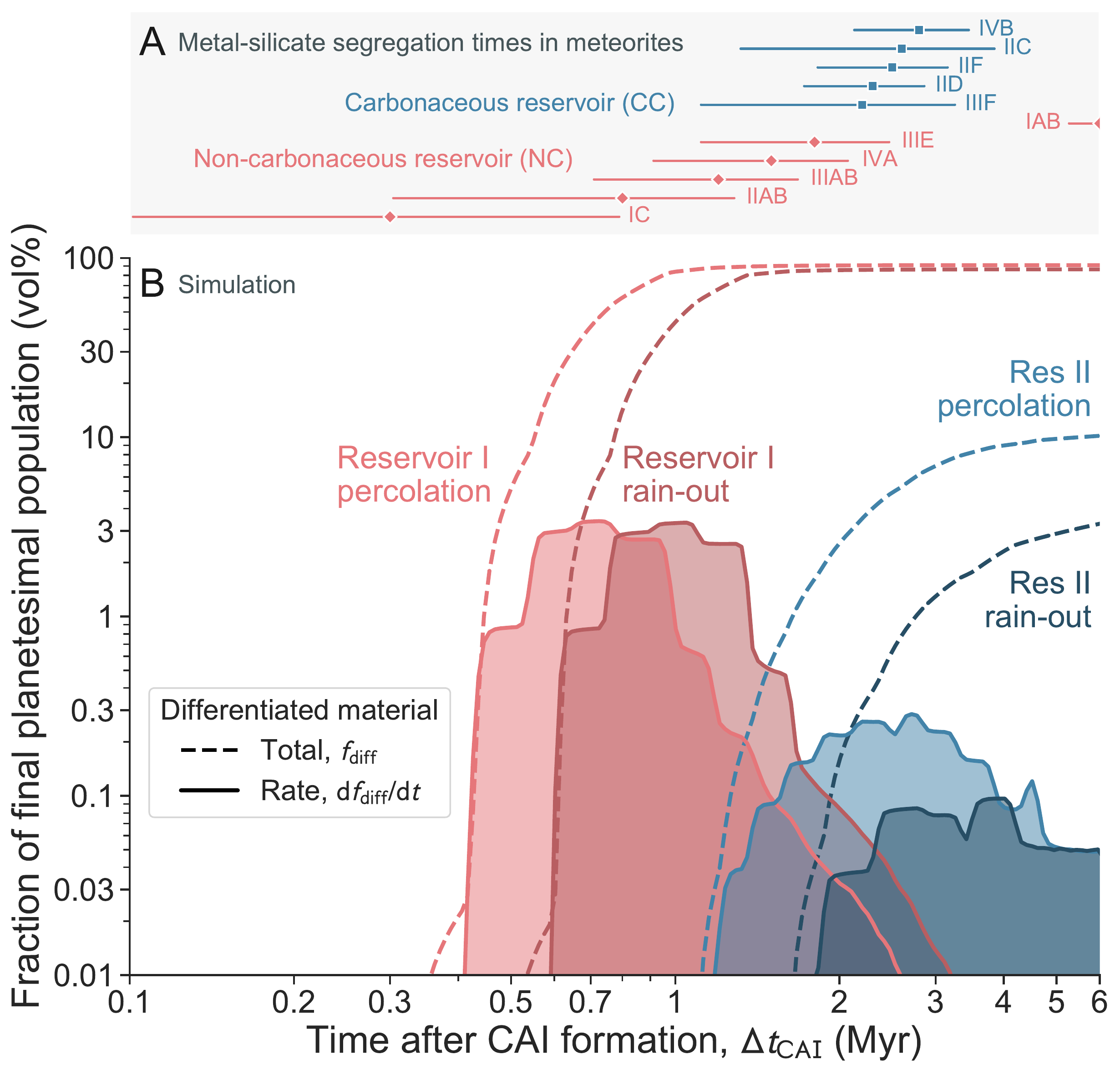}
   \caption{\textsf{\textbf{Comparison of metal-silicate separation times in simulated planetesimal populations with the meteorite record.} (\textbf{A}) Metal-silicate separation times in NC and CC meteorite classes (Tab.~S1). (\textbf{B}) Timing of core formation in Reservoir I (red) or Reservoir II (blue). Solid lines with colored shading represent the fraction of material undergoing metal-silicate separation (cf. Fig. \ref{fig:3} and \emph{Supplementary Materials}). Dashed lines show the total volumetric fraction for each scenario over time. Light and dark red/blue indicate percolation or rain-out, respectively.}}
   \label{fig:4}
\end{figure}

\begin{figure}[t]
   \centering
   \includegraphics[width=0.49\textwidth]{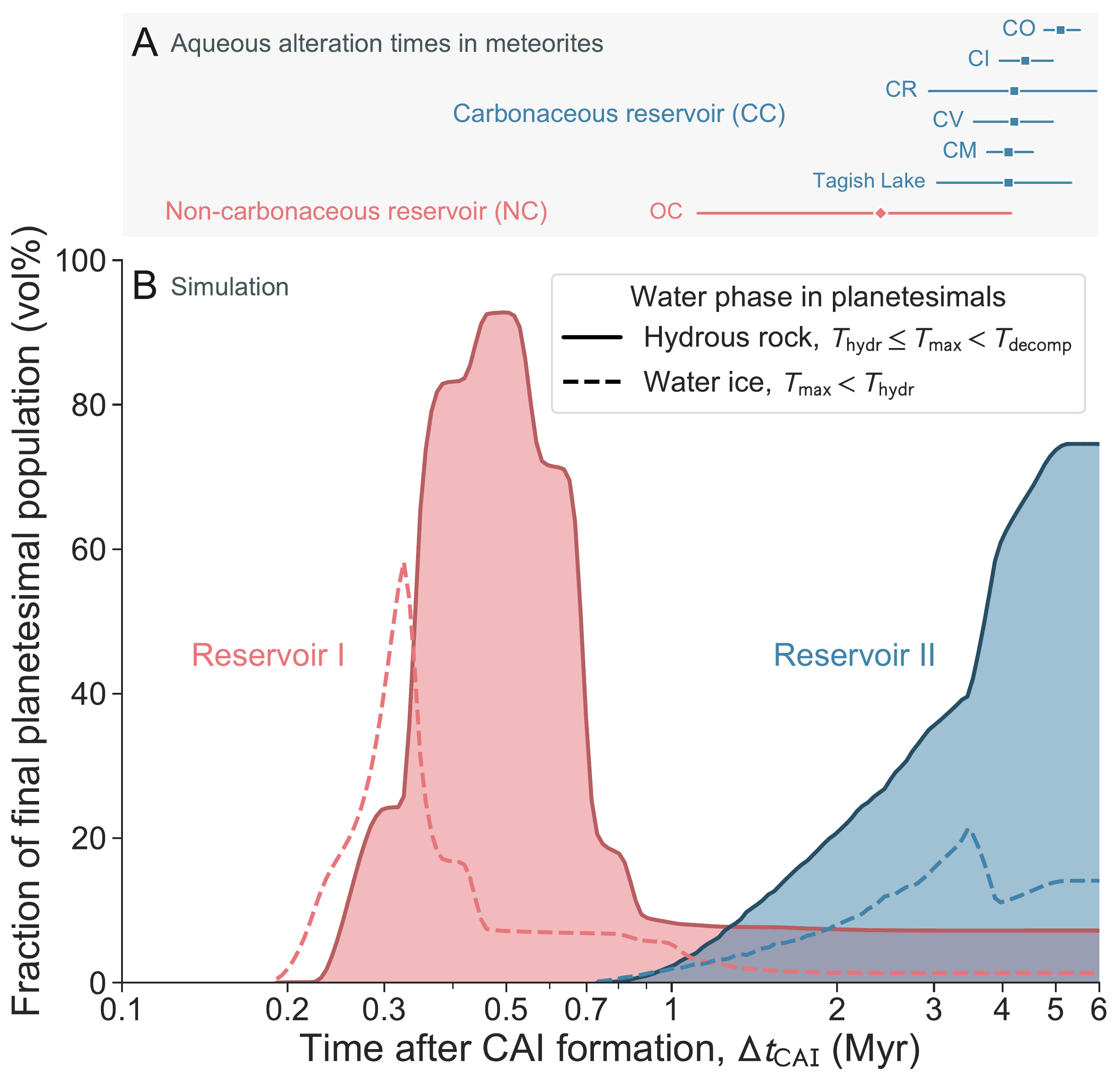}
   \caption{\textsf{\textbf{Comparison of hydrothermal activity in simulated planetesimal populations with aqueous alteration times in the meteorite record.} (\textbf{A}) Aqueous alteration times in NC and CC meteorite classes (Tab.~S1). (\textbf{B}) Fraction of planetary materials that retain primordial water ice (dashed lines) or undergo hydrothermal activity and retain water in hydrous silicate phases (solid lines with colored shading) in the simulations.}}
   \label{fig:5}
\end{figure}

Fig.~\ref{fig:4} compares the radiometric ages of metal-silicate separation in meteoritic measurements with the thermochemical evolution in our simulations. Reservoir I planetesimals undergo core formation between $\approx$0.4--3.0 Myr after CAI formation, with a peak between $\approx$0.6--1.3 Myr. On average, core formation via percolation occurs $\approx$0.4 Myr earlier than via rain-out in internal magma oceans. Multi-stage core formation scenarios, in which cores form by multiple stages of percolation and rain-out, are bracketed by these time intervals. Reservoir II planetesimals undergo core formation at later stages, starting at $\approx$1.1 Myr after CAI formation for percolation, and at $\approx$1.8 Myr after CAI formation for rain-out, and peaking between $\approx$1.8--3.1 and $\approx$2.3--4.0 Myr, respectively. The peak interval and the fraction of material experiencing differentiation in Reservoir I is earlier and higher than in Reservoir II due to the differing formation times and thus varying initial abundance of $^{26}$Al. The timing of core formation in both simulated populations agree with the meteorite data within the uncertainties of the individual ages \citep{2014Sci...344.1150K,2017PNAS..114.6712K,Hunt18EPSL}. The spread in early-formed NCs is reproduced within the uncertainties, with the exception of the IAB meteorite group \citep{Hunt18EPSL}. The fractional volumes in both simulated reservoirs differ, with $\approx$2--10 vol\% of Reservoir II undergoing core formation, whereas Reservoir I planetesimals differentiate to $\gtrsim$90 vol\%.

Fig.~\ref{fig:5} compares aqueous alteration ages from meteorites with our simulations. Simulated planetesimals in Reservoir I experience a brief phase of hydrothermal activity between $\approx$0.25--0.7~Myr after CAI formation and then dehydrate rapidly at $\approx$0.7 Myr. After that initial peak of water ice melting and hydrothermal activity, $\approx$10 vol\% of rock contains hydrous silicate phases, and $\lesssim$1 vol\% of water ice is retained. Reservoir II planetesimals, which form later with less $^{26}$Al, experience protracted hydrothermal activity lasting for several Myr with a peak at $\approx$5 Myr after CAI formation. At $\approx$3.5~Myr after CAI formation, more than 50~vol\% of the Reservoir II planetesimals have undergone hydrothermal activity. At $\approx$5~Myr after CAI formation, the simulation reaches a steady-state with $\approx$15~vol\% water ice and $\approx$75~vol\% hydrous rock. The peak for hydrothermal activity in Reservoir II reproduces the clustering of aqueous alteration in the CC meteorite record. The single available age for the NC population (an ordinary chondrite, OC) does not coincide with the peak in the Reservoir~I population, which we discuss below.
\section*{Compositional dichotomy between inner and outer Solar System} 
\label{sec:discussion_1}

\begin{figure*}[tbh!]
   \centering
   \includegraphics[width=0.99\textwidth]{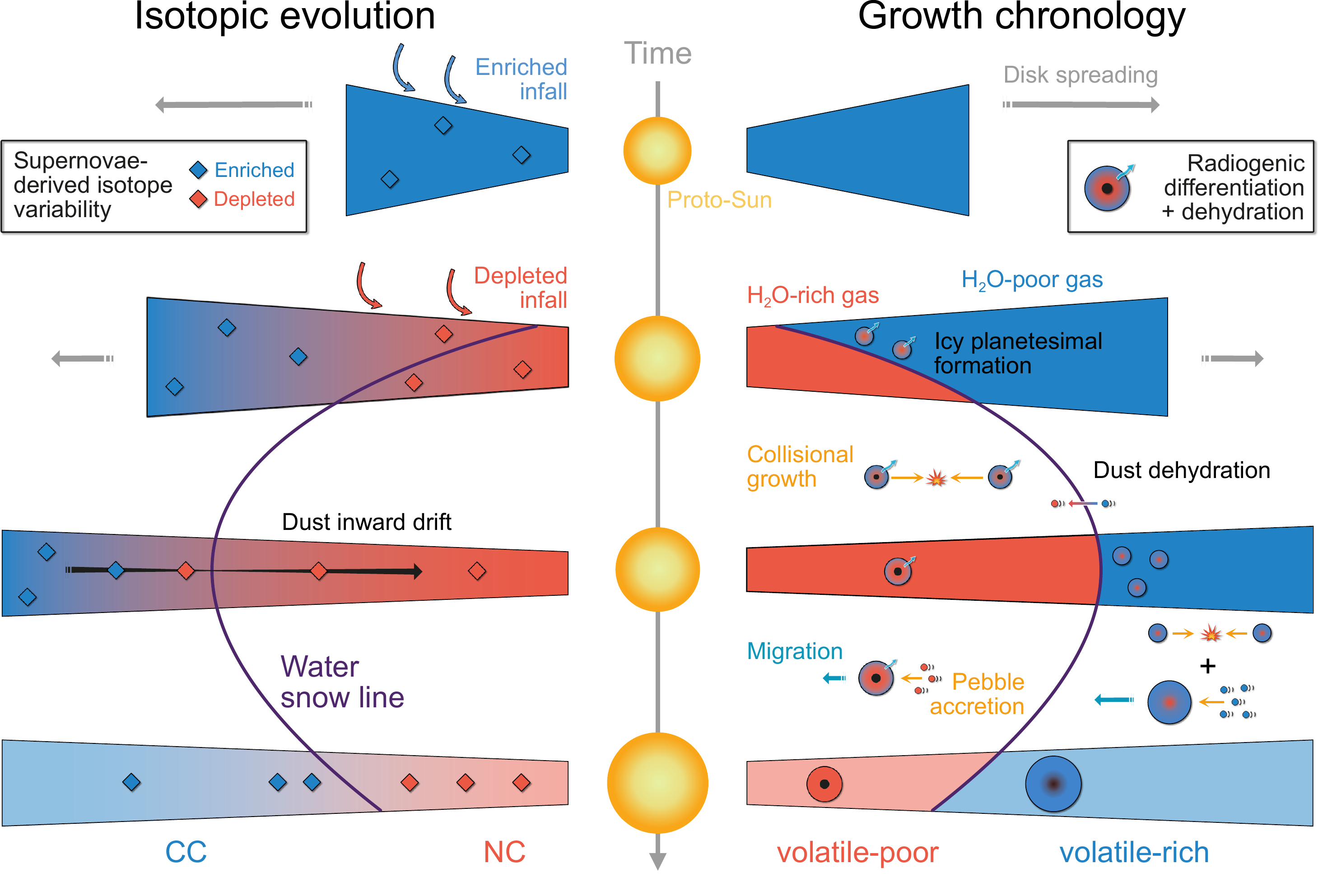}
   \caption{\textsf{\textbf{Schematic illustration of our proposed chronology of early Solar System accretion.} Nucleosynthetic isotope variability (left) across the disk due to varying composition of infall material is retained by the pile-up of inward-drifting dust grains at the snow line. The formation of two distinct planetesimal populations initiates divergent evolutionary pathways of inner and outer Solar System (right) due to the secular variation of local material composition, internal radiogenic heating, and dominant mode of planetary growth.}}
   \label{fig:6}
\end{figure*}

Our simulations indicate that the formation of spatially and temporally distinct planetesimal populations resulted in differing evolutionary pathways for planet formation in the inner and outer Solar System. The initial seed-population of the inner Solar System planets formed during the infall stage, while the outer Solar System planetesimal population started to form later, at the beginning of the Class II disk stage. Fig.~\ref{fig:6} shows our interpretation of the isotopic evolution during early Solar System formation. Measurements of CAIs indicate that the earliest Solar System infall material was initially dominated by dust enriched in supernovae-derived isotopes, then transitioned to a more depleted isotopic composition \citep{2018ApJS..238...11D,2019E&PSL.511...44N,2019ApJ...884...32J}. The distribution of isotopes evolved with time because of viscous disk spreading and increasing specific angular momentum of infalling matter. As a result, the outer disk became progressively dominated by early-infalling, enriched material, whereas the inner disk became dominated by late-infalling, depleted material \citep{2018ApJS..238...11D,2019E&PSL.511...44N,2019ApJ...884...32J}. We suggest that the distinct NC and CC reservoirs measured in the meteorite record result from the combined influence of early planetesimal formation in the inner Solar System and the subsequent dust pile-up at the snow line. The latter effect suppressed the rapid inward drift of enriched material and prevented isotopic homogenization of the disk. It also led to the onset of planetesimal formation and planetary accretion in the outer Solar System, which further limited mixing by dust drift from the outer to the inner reservoir. 

Fig.~\ref{fig:6} also illustrates our proposed timeline of compositional evolution and accretion, based on our simulations. The first planetesimal population formed due to the cold-finger effect. These bodies incorporated a substantial amount of $^{26}$Al and thus dehydrated and differentiated rapidly. Initial growth in the inner Solar System proceeded from mutual collisions among planetesimals to an intermediate phase of pebble accretion. During the final phase before disk dissipation, the pebble flux in the inner Solar System diminished and pebble growth may have stalled (see below). The outer Solar System population began to form later, during the Class II disk stage, and incorporated more total dust mass into its planetesimals. The initial phase of collisional growth was then succeeded by pebble accretion. This series of growth stages proceeded faster than in the inner Solar System due to the higher local pebble flux.

This chronology is consistent with meteoritic and astronomical evidence and makes several potentially testable predictions. The rapid decrease in the pebble flux in the inner Solar System (Reservoir I), to less than one Earth mass per Myr, reduces the growth by pebble accretion during later disk stages. Parameter space exploration (\emph{Supplementary Materials}) and previous work \citep{2015PNAS..11214180L} indicate that under such conditions pebble accretion onto inner Solar System planetary embryos is inefficient. This implies that the intermediate, pebble-driven accretion phase stalled, and subsequent growth of the terrestrial protoplanets was driven by collisional interactions during the final disk stages and afterwards. Such protracted accretion timescales for the terrestrial planets are supported by radiometric evidence for both proto-Earth and proto-Mars \citep{2011Natur.473..489D,2010NatGe...3..439R}, and the absence of rocky planets larger than Earth (super-Earths) in the inner Solar System. The formation of super-Earth exoplanets has been attributed to a high pebble flux and rapid inward migration with increasing planet mass \citep[][\emph{Supplementary Materials}]{Raymond2020PlanetaryAstrobiology,2019A&A...624A.109B,2019A&A...624A.114L}. The secular transition of an early, low pebble flux to a brief period of high pebble flux could amplify preceding mass differences between accreting protoplanets and thus explain why Earth is the inner Solar System body with the closest composition to CI-chondrites \citep{2009Sci...324..374T,2010Sci...328..884S,2018Natur.555..507S,2019NatGe..12..564B}.

The secular transition between growth regimes, the internal evolution of planetesimals, and the temporal variation of the local pebble composition would lead to a heterogeneous delivery of water and other highly volatile compounds to the inner Solar System (\emph{Supplementary Materials}): the birth planetesimals in Reservoir I form an initial seed-population of water-rich bodies that subsequently dehydrate from $^{26}$Al heating and accrete via collisions. With progressing time and planetary growth, the protoplanets in the inner Solar System experience an influx of dry pebbles between $\approx$1--2 Myr after CAI formation. Forming such initially icy and subsequently dehydrating planetesimals in the inner Solar System is consistent with measurements of hydrogen content in enstatite \citep{Piani+2020}, carbonaceous chondrite \citep{2018NatAs...2..317P}, eucrite and angrite \citep{2017RSPTA.37560209S} meteorites, Earth's deep mantle, and bulk measurements of the terrestrial planets and asteroid families \citep{PeslierISSI2018,2018SSRv..214...36A}. The timing of hydrothermal activity in Reservoir I planetesimals in our simulations, at $\approx$0.3--0.7 Myr after CAI formation, is offset from the Reservoir II peak at $\gtrsim$4 Myr after CAI formation (Fig. \ref{fig:5}). Most of the hydrous phases that are formed in the Reservoir I peak, however, are subsequently destroyed by high-temperature internal processing in our simulations, consistent with evidence from NC meteorites \citep{2016M&PS...51.1886L}. Such intense (planetary) thermal processing in Reservoir I may thus lead to fractionation in mass-dependent isotopes \citep{2020Icar..34713772B}, consistent with  evidence from terrestrial planetary bodies \citep{2017Natur.549..511H,2017Natur.549..507N}. The disk simulation does not produce direct analogues of ordinary or enstatite chondrites from gravitational collapse, but the predicted growth stages (Figs. \ref{fig:2} and \ref{fig:6}) would lead to recycling of collisionally disrupted planetesimals \citep{2018Icar..302...27L} and secondary layers accreted onto growing bodies \citep{2020SciA....6A1303M}, which could be potential origin locations.
The formation of Reservoir II planetesimals with initially higher-than-today water abundances is consistent with evidence for extensive iron oxidation and deuterium enrichment during aqueous alteration of CC meteorites \citep{2017GeCoA.211..115S}. Simulated Reservoir II planetesimals undergo a prolonged Myr-long phase of hydrothermal activity with a peak at $\approx$4--5 Myr after CAI formation, which reproduces the aqueous alteration ages of CC meteorites \citep{2018SSRv..214...36A}. However, later-formed Reservoir II bodies remain almost unmodified by hydrothermal alteration (Fig. \ref{fig:3}), so surviving bodies today should display varying internal processing, ranging from extensive aqueous alteration to fully pristine for small and late-formed bodies. Reservoir II planetesimals sample material originating from a wide range of orbital distances and form over an extended time interval. Therefore, compared to Reservoir I planetary materials, Reservoir II bodies may exhibit larger intra-reservoir isotope variability. This is consistent  with the nucleosynthetic isotope variability \citep{2008E&PSL.266..233L} and abundance of presolar interstellar materials \citep{2014GeCoA.139..248D} in carbonaceous chondrites, and the scatter in cometary D/H ratios \citep{2018SSRv..214...36A}.

In summary, the chronology of Solar System formation (Fig.~\ref{fig:6}) we infer from our simulations links several characteristics found by geochemical laboratory analyses and astronomical observations. We interpret the chemical (volatile-poor versus volatile-rich) and isotopic (NC versus CC) dichotomy as causally linked by the build-up of distinct planetesimal populations in the inner and outer Solar System. Mixing by dust drift between them is limited to the earliest disk phases and declining with time due to the progressive accretion of the outer Solar System planetary population. The temporal and spatial variation in the main volatile reservoirs and heterogeneous planet growth are intrinsically coupled to the disk structure, the redistribution and fractionation of volatile ices, the abundance of short-lived radionuclides, and the geophysical evolution of protoplanets during planetary formation.

\paragraph{Acknowledgements}
We thank C. P. Dullemond, A. C. Hunt, I. Pascucci, S. Ida, J. Wade, S.-J. Paardekooper, T. Birnstiel, S. M. Stammler, J. J. Barnes, A. Morbidelli, W. Kley, and members of the ERC EXOCONDENSE project at Oxford for discussions; B. Liu (\begin{CJK*}{UTF8}{gbsn}刘倍贝\end{CJK*}\hspace{-0.1cm}), J. F. J. Bryson, M. Ek, R. D. Alexander, and S. Charnoz for comments on earlier draft versions; T. V. Gerya for usage of the \texttt{I2ELVIS} code family; and C. P. Dullemond for usage of the disk evolution code.
\textsc{Funding:}  T.L. received funding from the Simons Foundation (SCOL award \#611576) and the Swiss National Science Foundation (grant \#P2EZP2-178621). J.D. received funding from the European Research Council under the European Union’s Horizon 2020 research and innovation program (grant \#714769). T.O.H. was supported by the University of Zurich Forschungskredit Postdoc. Parts of this work were carried out within the framework of the National Centre for Competence in Research PlanetS (grant \#51NF40-141881) supported by the Swiss National Science Foundation.
\textsc{Author contributions:} 
Conceptualization, T.L., J.D., M.S., G.J.G.; 
Methodology, T.L., J.D., T.O.H., G.J.G.;
Software, T.L., J.D., T.O.H., G.J.G.;
Validation, all authors;
Formal Analysis, T.L., J.D., T.O.H.; 
Investigation, T.L., J.D., M.S.;
Resources, T.L., J.D., M.S.;
Writing, T.L., J.D., T.O.H., M.S.;
Review \& Editing, all authors;
Visualization, T.L.; 
\textsc{Competing interests:} We declare no competing interests.
\textsc{Data and materials availability:} Simulation codes and output data are available at \href{https://osf.io/e2kfv/}{osf.io/e2kfv}.

\balance


\vfill 
\pagebreak
\FloatBarrier

\section*{\large{Supplementary Materials}}
\label{sec:suppl}

\renewcommand{\thefigure}{S\arabic{figure}}
\renewcommand{\thetable}{S\arabic{table}}
\renewcommand{\thepage}{S\arabic{page}}
\renewcommand{\theequation}{S\arabic{equation}}
\setcounter{figure}{0}
\setcounter{page}{1}

\subsection*{Materials \& Methods}
\label{sec:methods}

\subsubsection*{Disk evolution \& planetesimal formation}
\label{sec:disk_evolution}

Planetesimals, the gravitationally bound seeds of the accretion process and building blocks of planets, are thought to form in a multi-stage process with dust first growing to pebble sizes, which can then be concentrated by the streaming instability \citep{2014prpl.conf..547J}. The streaming instability is a two-fluid mechanism driven by the relative flow of gas and dust which leads to the formation of dense dust filaments in circumstellar disks. Under certain conditions, those filaments may become massive enough to collapse under their own gravity, leading to rapid planetesimal formation \citep{2005ApJ...620..459Y,2007Natur.448.1022J}. Planetesimals may be formed throughout the protoplanetary disk as long as both gas and pebbles are present to create localized overdensities that breach the physical conditions for gravitational collapse. To calculate the timing and integrated mass of planetesimals formed via the streaming instability, we consider a physical model of the nebular gas disk, dust evolution including growth and fragmentation \citep{2016SSRv..205...41B}, and a prescription for turning pebbles into planetesimals that parameterizes the effects of gravitational collapse and planetesimal formation as obtained from fluid dynamical models.

\begin{figure}[th]
   \centering
   \includegraphics[width=0.49\textwidth]{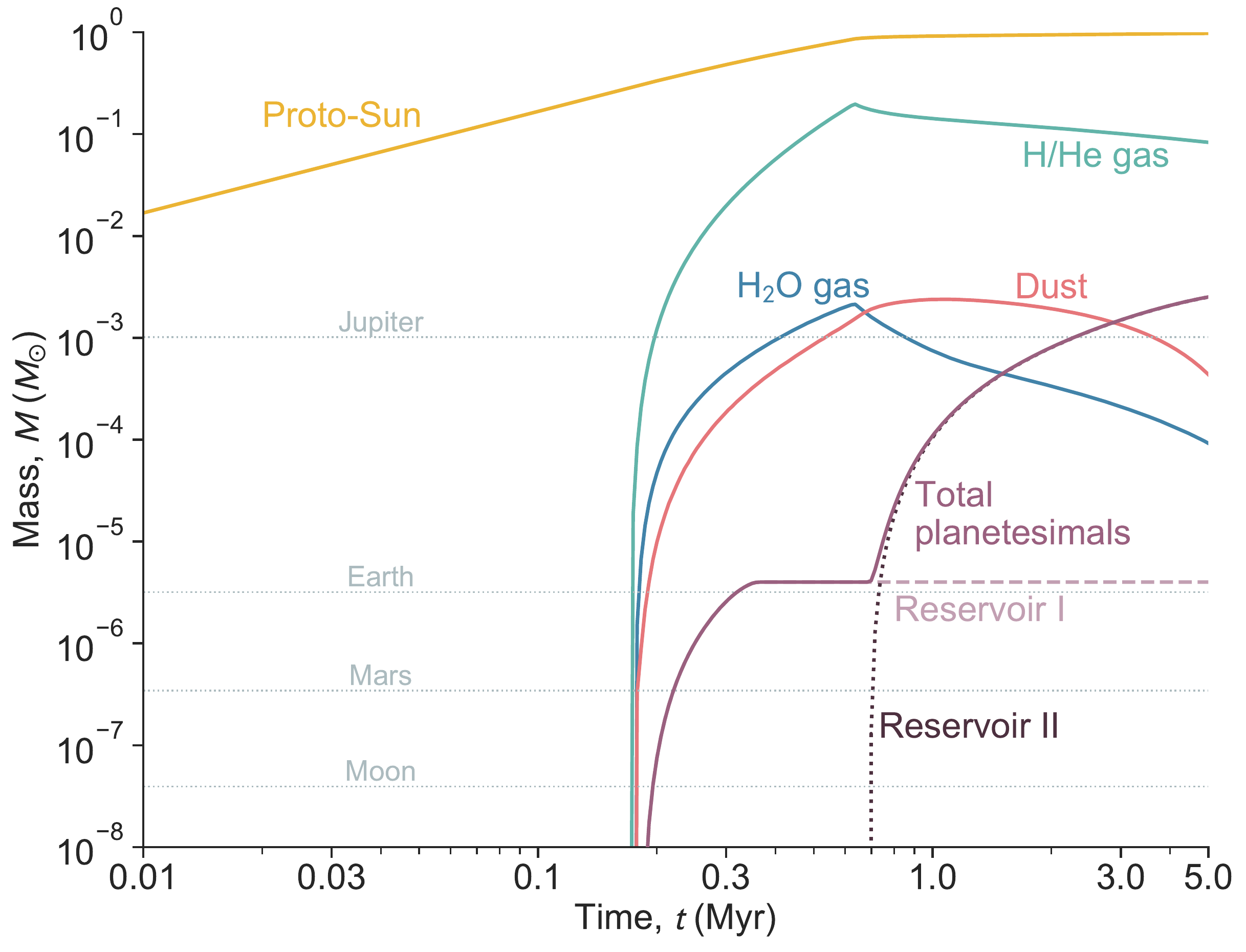}
   \caption{\textsf{\textbf{Mass evolution for different reservoirs in the disk simulation} \citep{DD18}. The mass of the proto-Sun (yellow line), H/He gas (green), H$_2$O gas (blue), and dust (red) grow from the surrounding molecular cloud until the end of the Class I disk stage at $\approx$0.7~Myr. The mass in planetesimals (solid purple line) is divided into Reservoir I (dashed purple line) and Reservoir II (dotted purple line) as defined in Fig.~\ref{fig:1}. Overplotted are the present-day masses of several Solar System objects (horizontal lines).}}
   \label{fig:s1}
\end{figure}

\begin{figure*}[tbh]
   \centering
   \includegraphics[width=0.99\textwidth]{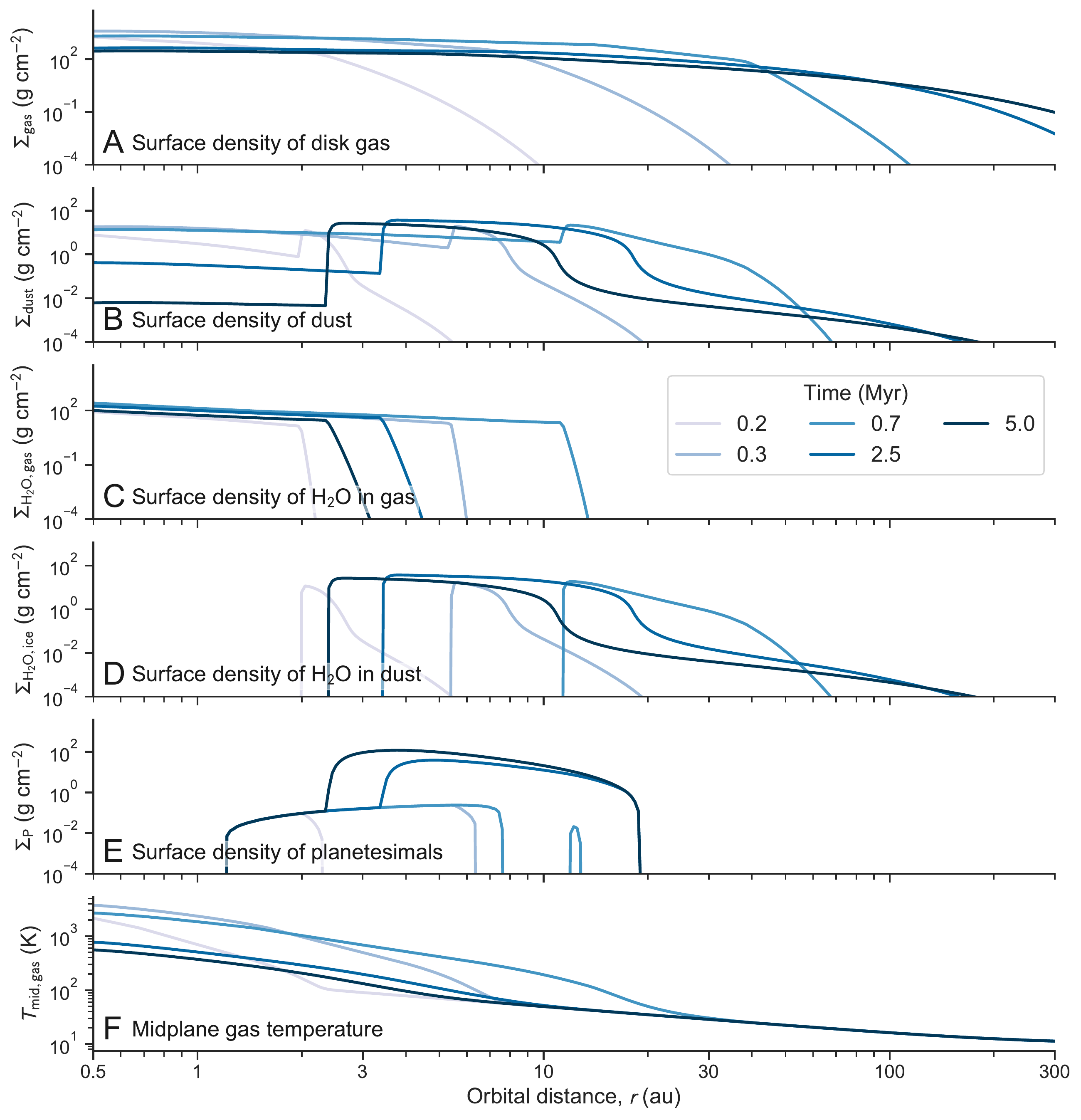}
   \caption{\textsf{\textbf{Radial distribution and time evolution of the composition in the disk simulation} \citep{DD18}. Shown are the surface densities of H/He gas (A), dust (B), H$_2$O gas (C), H$_2$O ice (D), planetesimals (E), and midplane gas temperature at 0.2, 0.3, 0.7, 2.5, and 5.0 Myr (light to dark blue).}}
   \label{fig:s2}
\end{figure*}

We make use of previously-published simulations \citep{DD18} that include a one dimensional model in which the protoplanetary disk is formed from the parent molecular cloud and undergoes viscous evolution. These simulations follow the rotating infalling cloud model \citep{2005A&A...442..703H, 2006ApJ...640L..67D, 2006ApJ...645L..69D}. In this model there are essentially three components: the parent molecular cloud, which rotates as a solid body, the central star, and the disk, both of which are growing, fed by the infalling cloud. We use this simulation to follow the gas and dust evolution from the infall (Class I) to the Class II disk stage, for which there is observational evidence for declining dust mass with time \citep{2016ApJ...828...46A,2016ApJ...831..125P,2018ApJ...869L..41A,2020A&A...640A..19T}. The disk simulation assumes an initial mass of 1 solar mass (M$_\odot$), an isotropic temperature of 10~K, and a rotation rate of $5 \times 10^{-15}$~s$^{-1}$ for the parent molecular cloud of the Solar System. After about $7\times10^{5}$~yr this cloud forms a single central star surrounded by a circumstellar disk with a peak total mass of 0.2~M$_{\odot}$ (Fig. \ref{fig:s1}).
The temperature of the disk is calculated taking into account heating due to viscosity and irradiation by the central star. The evolution of the gas surface  density $\Sigma_{\rm g}$ (Fig. \ref{fig:s2}) is computed using
\begin{equation}\label{gaseq}
\frac{\partial \Sigma_{\rm g}}{\partial t} + \frac{1}{r} \frac{\partial}{\partial r}\left({r\Sigma_{\rm g}v_r}\right) = S_{\rm g},
\end{equation}
where $r$ denotes the radial distance to the central star, $t$ the time, $v_r$ is the radial velocity of the gas, and $S_{\rm g}$ represents a source term due to matter infalling onto the disk from the molecular cloud. The viscous evolution of gas follows the standard $\alpha$-formalism \citep{1973A&A....24..337S}, where the gas viscosity is defined as
\begin{equation}\label{eq:gasvis}
\nu = \alpha_{\rm v} c_{\rm s} H_{\rm g},
\end{equation}
where $c_{\rm s}$ is the sound speed, $H_{\rm g}$ is the gas scale height, and $\alpha_{\rm v}$ is the disk viscosity, parameterizing the physical mechanism of angular momentum transport of the disk. The viscosity is fixed to $\alpha_{\rm v}=10^{-3}$. We refer to \citep{DD18} for the effects of a varying $\alpha_{\rm v}$ on dust coagulation and planetesimal formation. The viscosity determines the radial gas velocity, which is calculated from
\begin{equation}\label{eq:gasvel}
v_r = -\frac{3}{\Sigma_{\rm g}\sqrt{r}}\frac{\partial}{\partial r}\left(\Sigma_{\rm g} \nu \sqrt{r}\right).
\end{equation}

\begin{figure*}[tbh]
   \centering
   \includegraphics[width=0.99\textwidth]{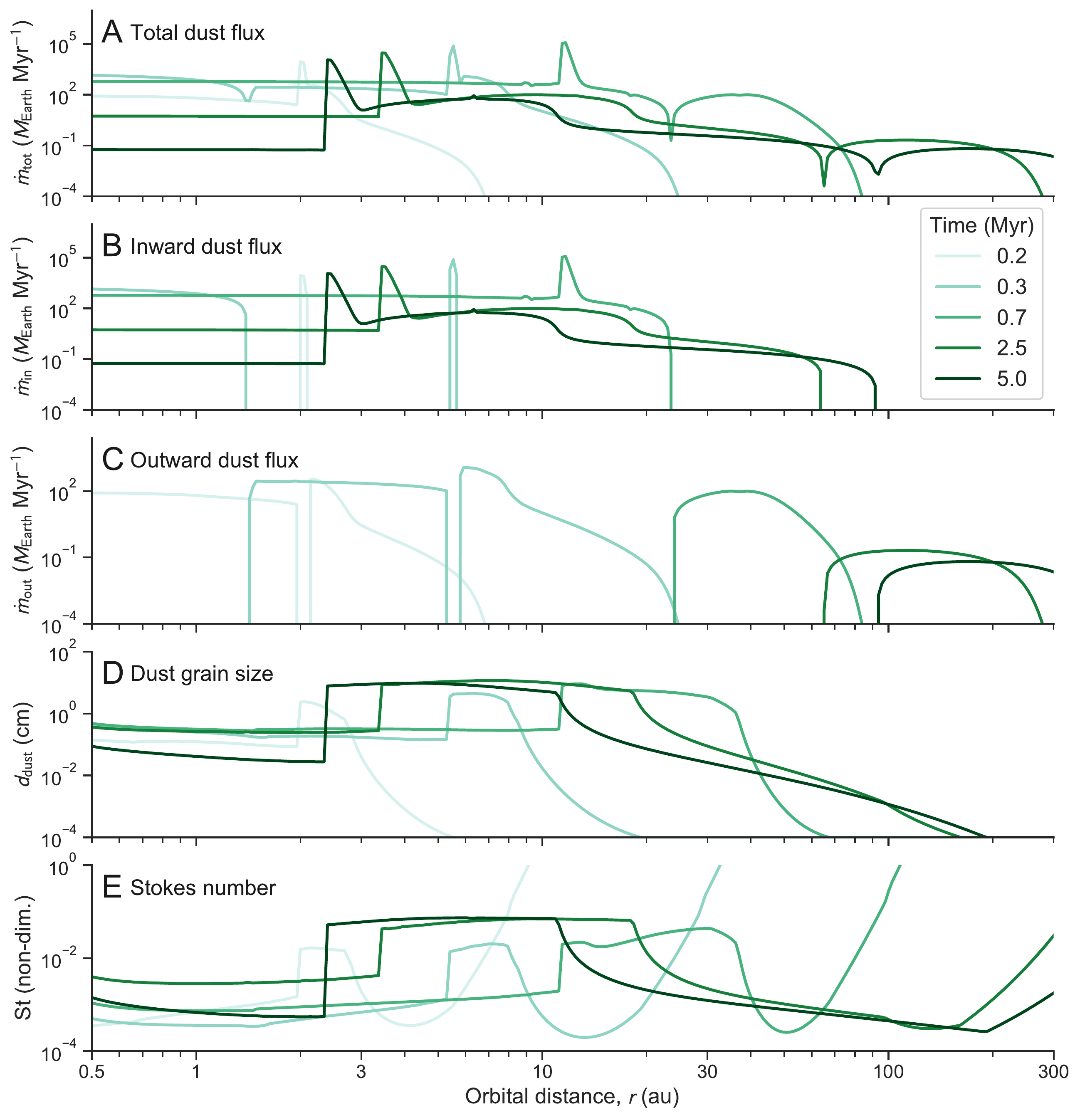}
   \caption{\textsf{\textbf{Radial distribution and time evolution of additional dust parameters in the disk simulation} \citep{DD18}. Shown are total dust flux (A), inward dust flux (B), outward dust flux (C), dust grain size (D), and the Stokes number (E) at 0.2, 0.3, 0.7, 2.5, and 5.0 Myr (light to dark green).}}
   \label{fig:s3}
\end{figure*}

The infall of gas is complemented by the delivery of dust (Fig. \ref{fig:s3}), with a dust-to-gas ratio of 1\%, as found in the interstellar medium \citep{2001RvMP...73.1031F}. We follow the evolution of dust surface density $\Sigma_{\rm d}$ using the advection-diffusion equation,
\begin{equation}\label{dusteq}
\frac{\partial \Sigma_{\rm d}}{\partial t} + \frac{1}{r} \frac{\partial}{\partial r}\left[r\left(\Sigma_{\rm d}\bar{v}-\nu\Sigma_{\rm g}\frac{\partial}{\partial r}\left[\frac{\Sigma_{\rm d}}{\Sigma_{\rm g}}\right]\right)\right] = S_{\rm d},
\end{equation}
where $S_{\rm d}$ is the source term for the $\mu$m-sized dust infalling from the molecular cloud and $\bar{v}$ is the mass-weighted averaged radial speed of dust grains, which depends on dust size. 
Dust and gas are assumed to be initially well-mixed. The dust grains start to decouple from the gas as they grow into larger (pebble) sizes. Dust grain growth is computed adopting an algorithm \citep{2012A&A...539A.148B} that relies on only tracking two populations of dust grains: the small and the large ones. Dust size is regulated by the initial growth phase and halted either by fragmentation or radial drift. 

Alongside dust growth, ice sublimation and water vapor re-condensation are incorporated \citep{2006Icar..181..178C}. The infalling dust is assumed to consist of 50\% refractory material and 50\% water ice. When the ice sublimates, its mass is added to the water vapor reservoir. For the gas disk content, the surface density of water vapour and hydrogen-helium is tracked separately. When the water vapor exceeds the equilibrium vapor pressure value, it condenses onto the existing grains, adding to their water ice content. In contrast, when the water vapor pressure is lower than the equilibrium pressure, sublimation takes place. This way, the position of the snow line can be measured as the location where the ice content of grains changes rapidly. The composition of grains alters their sticking properties such that ice-rich grains are more sticky compared to refractory, ice-free grains \citep{2009ApJ...702.1490W, 2011ApJ...737...36W, 2014MNRAS.437..690A}. Thus, the fragmentation threshold velocity for dry aggregates inside of the snow line is set to $v_{\mathrm{frag,in}}=1$~m s$^{-1}$ and for icy aggregates outside of the snow line to $v_{\mathrm{frag,out}}=10$~m s$^{-1}$.

Planetesimal formation is included by assuming that planetesimals form through the streaming instability when the midplane dust-to-gas ratio of pebbles, which are characterized by the Stokes number St $\geq$ 10$^{-2}$, exceeds unity. Whenever this criterion is fulfilled, part of the surface density of pebbles is transferred into planetesimals. For this to take place, a dense midplane layer of sufficiently large pebbles is required. In the disk model, these large pebbles only grow outside of the water snow line, where the dust is more sticky. Enhancements of the vertically integrated dust-to-gas ratio to trigger planetesimal formation have been shown to build-up in the inner part of the protoplanetary disk during long-term dust evolution and mass influx from the outer parts of the disk \citep{2016A&A...594A.105D}. The adopted disk model does not include a physical mechanism for gas disk dispersal, such as internal or external photoevaporation \citep{2017RSOS....470114E}, which can affect the total mass of planetesimals that are formed during the final stages of disk evolution. However, geochemical \citep{2017SciA....3E0407B} and paleomagnetic \citep{2017Sci...355..623W} proxies indicate a dispersal of the solar nebula on the order of 4--5 Myr, which is when we assume the disk to disperse and planetesimal formation to cease.

We focus on the physical processes taking place around the snow line, which facilitates dust-to-gas overdensities as the dry dust inside of the snow line reaches smaller sizes and thus drifts more slowly than the icy dust outside, which leads to a pile-up effect \citep{2016A&A...596L...3I, 2017A&A...608A..92D, 2017A&A...602A..21S}. However, the build-up of this enhancement similarly requires mass transfer from the outer to the inner disk on a timescale of $\gtrsim 10^{5}$~yr, depending on the assumed disk parameters. Thus, planetesimal formation triggered by this mechanism typically takes place in the Class II disk stage \citep{DD18}. An additional mechanism that operates on shorter timescales, but also leads to less pronounced enhancement of dust density, is the cold-finger effect \citep{1988Icar...75..146S,2004ApJ...614..490C}. This relies on the outward diffusion of water vapor from inside and its subsequent re-condensation onto the grains outside of the snow line. This mechanism produces a moderate enhancement of the vertically integrated dust-to-gas ratio during the protoplanetary disk build-up stage, and depends on the level of midplane gas turbulence.

In the adopted inside-out infall model, the snow line first moves outward during the infall phase, reaching its furthest location when the disk obtains its peak mass (at about 0.7~Myr) resulting from an increase in stellar luminosity and gas density during disk build-up (Fig. \ref{fig:s2}A). The temperature is calculated taking into account both the irradiation from the central star and dissipation of viscous energy by turbulence. Stellar irradiation sets a shallow temperature profile in the whole disk, while viscous heating increases the temperature dominantly in the inner parts, which leads to a steeper temperature profile close to the star compared to a profile with only irradiation (Fig. \ref{fig:s2}F). In the Class II phase, snow line movement is driven both by the evolution of the slowly cooling disk, and by dust evolution. The position of the snow line is self-consistently computed as the location at which the dust grains contain less than 1\% ice component, thus the snow line evolution is an outcome rather than an assumption of the simulation. The pebble pile-up outside of the snow line (Figs. \ref{fig:s2}B and D) increases the time it takes the solid ice component of pebbles to evaporate while they drift inward. As a result, the snow line position moves inward faster than it would just from the temperature evolution.

The position of the snow line is related to the temperature structure, which in the inner disk is dominated by the dissipation of viscous energy by turbulence. From astronomical observations of other protoplanetary disks, the level of turbulence, and thus the major physical mechanism of angular momentum transport, remains unclear. Some observational attempts to measure turbulence have been inconclusive \citep{2015ApJ...813...99F, 2016A&A...592A..49T}. In a standard model of protoplanetary disk evolution \citep{1991ApJ...376..214B}, the angular momentum transfer is driven by isotropic turbulence, such that the ratio of midplane turbulence ($\alpha_{\rm t}$) equals the disk viscosity, hence $\alpha_{\rm v}=\alpha_{\rm t}$. However, large regions of the protoplanetary disk, mostly surrounding its midplane, are expected to be free from turbulence \citep{1996ApJ...457..355G}. This is supported by observational evidence for turbulence levels that are lower than anticipated \citep{2017ApJ...843..150F,2019Natur.574..378T,2020ApJ...895..109F}. The disk model we employ \citep{DD18}  mimics a disk structure in which the midplane turbulence is lower than that expected from the global angular momentum transport rate by decoupling the ratio of midplane turbulence, $\alpha_{\rm t}$, from the global viscous $\alpha$-parameter, $\alpha_{\rm v}$. This is a simple mimicking of a potentially much more complex structure of a non-ideal magneto-hydrodynamic disk, in which large regions of the midplane are free from turbulence and the angular momentum can be removed vertically by magnetic winds or radially by laminar torques \citep{2014A&A...566A..56L, 2016ApJ...818..152B, 2017ApJ...845...75B}. Similar approaches were used in other recent models \citep{2017ApJ...839...16C, 2017MNRAS.472.4117E}. A parameter study of $\alpha_{\rm v}$ and $\alpha_{\rm t}$ space has been performed for this model \citep{DD18}, which concluded that early planetesimal formation in the infall stage is only possible if $\alpha_{\rm t} \ll \alpha_{\rm v}$ because planetesimal formation from the cold-finger effect is favoured by near-laminar midplane conditions. We focus on the scenario with $\alpha_{\rm t} = 10^{-5}$ and $\alpha_{\rm v} = 10^{-3}$ \citep{DD18}. However, our results would be qualitatively similar for any model forming planetesimals both during the infall and the disk stage.

Decoupling the $\alpha_{\rm v}$ and $\alpha_{\rm t}$ values aims to mimic the layered accretion scenario in which the gas flow takes place in active layers above a laminar midplane. However, the disk model is one dimensional and vertically integrated. Thus, $\alpha_{\rm v}$ describes the density-averaged gas flow and is used to calculate the disk temperature in the viscous heating regime. This leads to intermittent high temperatures in the inner disk and pushes the water ice line to radii beyond 10 au for a short time interval. Calculating the midplane temperature is generally a complex radiative transfer problem \citep{2014prpl.conf..411T} and numerical models vary in their conclusions on the midplane temperature in the inner parts of the disk \citep{2013A&A...560A..43F,2019ApJ...881...56S}, associated dust redistribution and composition, and hence planetesimal formation rates \citep{2016A&A...589A..15S,2019A&A...627A..50C,2019A&A...624A..28I}.

\subsubsection*{thermochemical evolution of planetesimals}
\label{sec:planetesimal_evolution}

\begin{figure*}[tbh]
   \centering
   \includegraphics[width=0.49\textwidth]{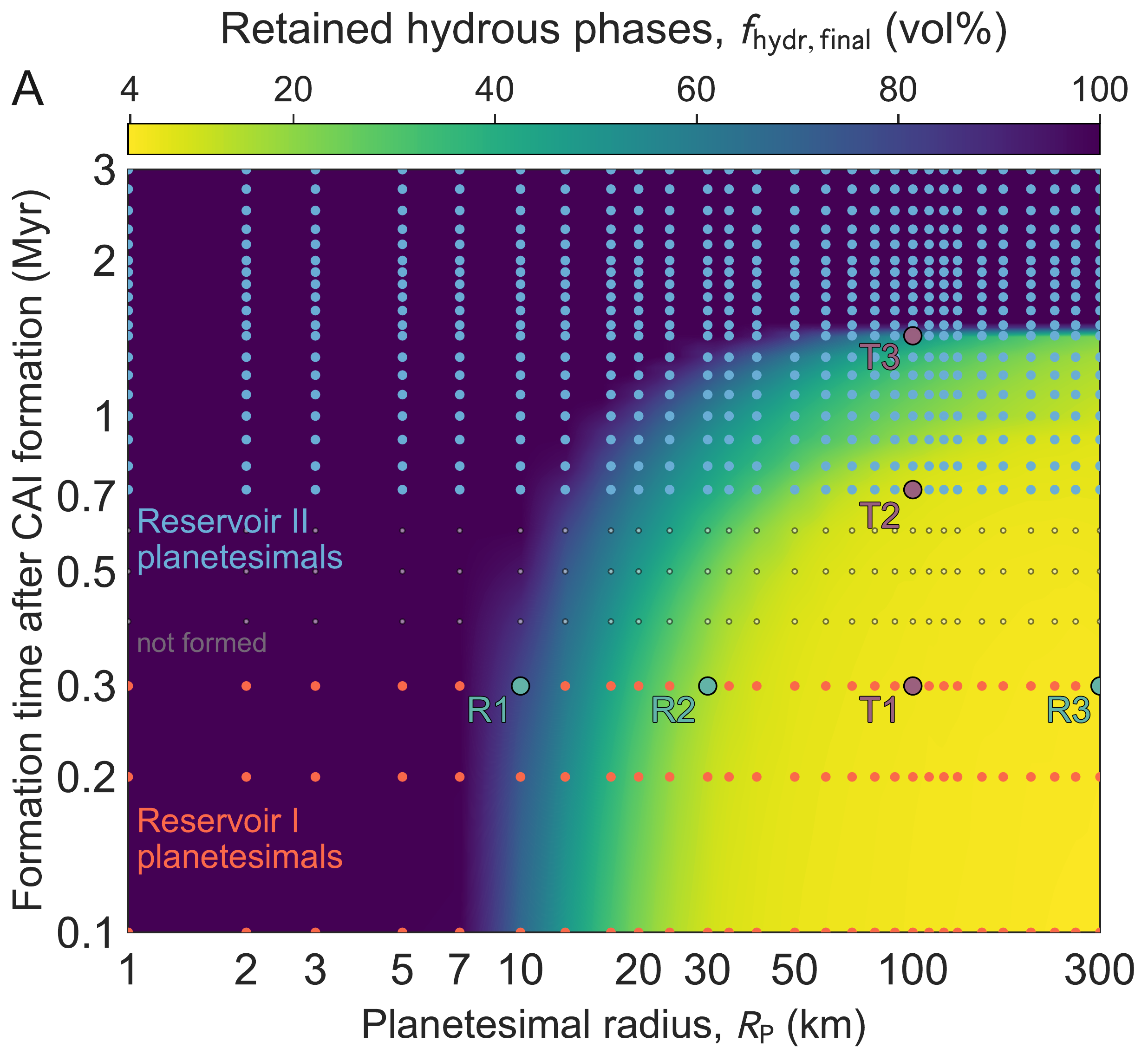}
   \includegraphics[width=0.49\textwidth]{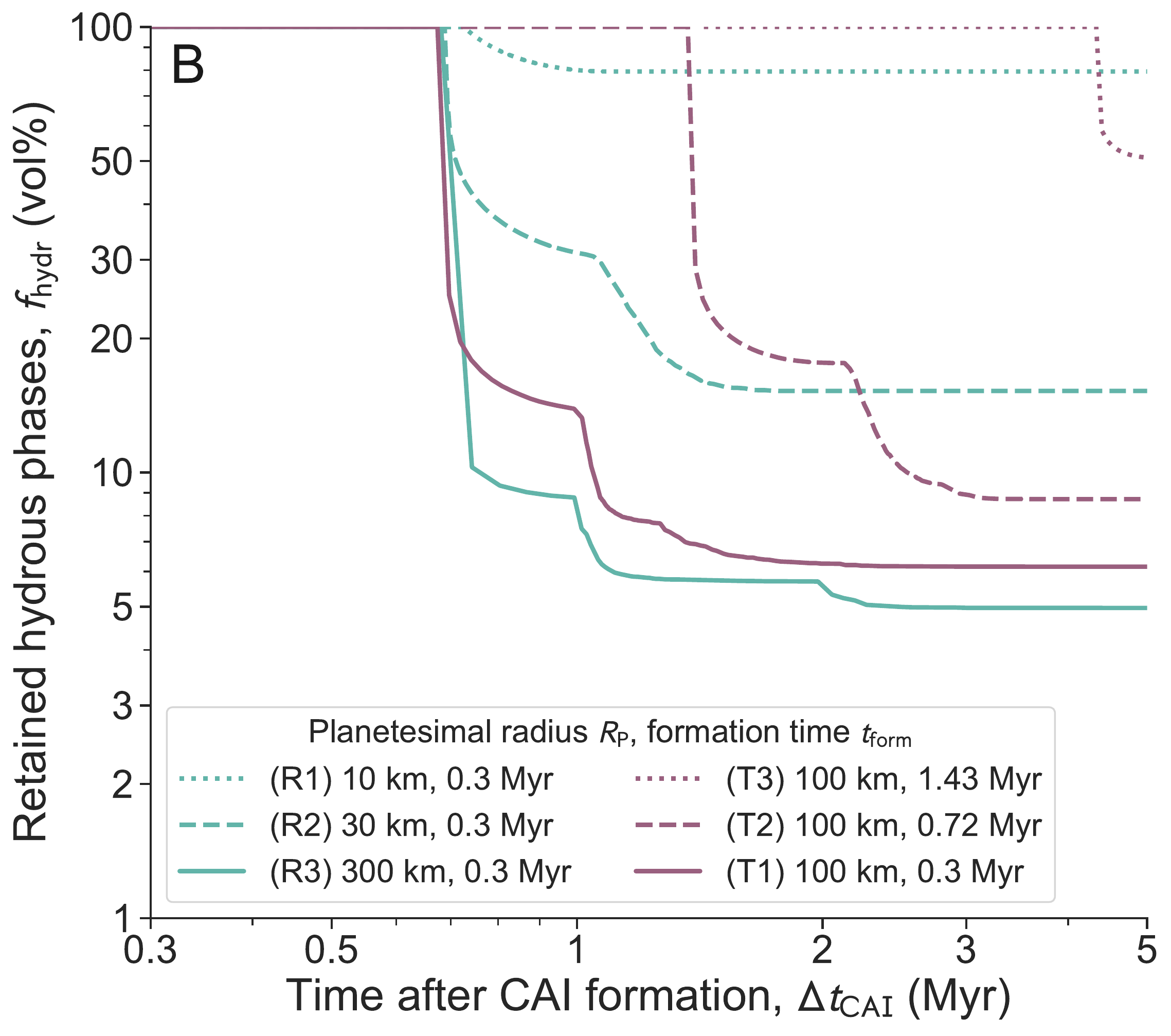}
   \caption{\textsf{\textbf{Parameter space and time evolution of retained hydrous rock phases in planetesimal simulations}. (\textbf{A}) Analogous to Fig.~\ref{fig:3}A, but for the retained volumetric fraction of hydrous rock phases in simulated planetesimals after 5 Myr (thermochemical criterion $T < T_\mathrm{decomp}$). Gray dots (labeled 'not formed') are used for the color interpolation, but do not contribute to Reservoir I or II. (\textbf{B}) Time evolution for distinct combinations of formation time and planetesimal radius. Lines R1--R3 and T1--T3 indicate the evolution of planetesimal simulations labeled in panel \textbf{A}.}}
   \label{fig:s4}
\end{figure*}
\begin{figure*}[tbh]
   \centering
   \includegraphics[width=0.49\textwidth]{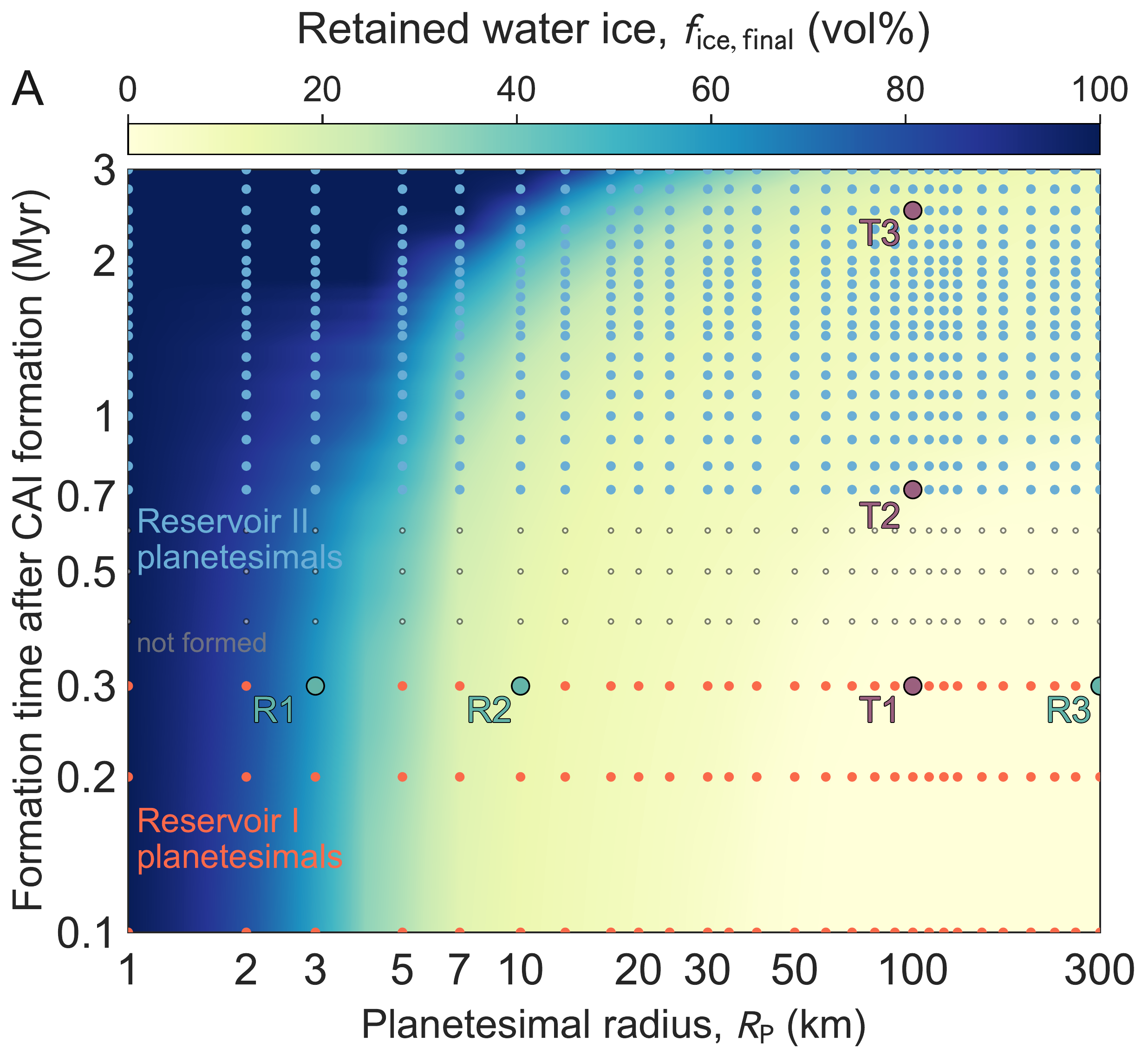}
   \includegraphics[width=0.49\textwidth]{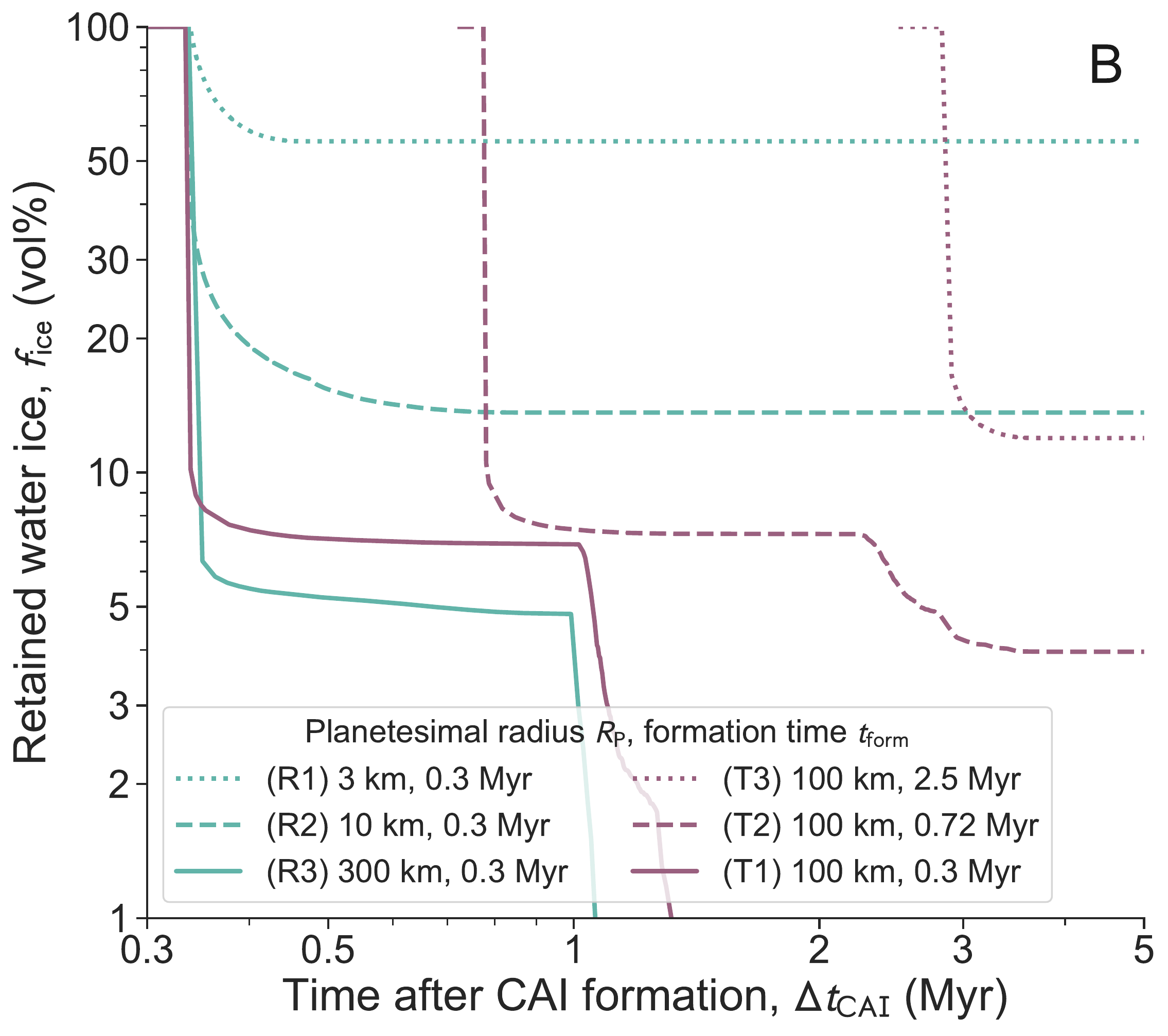}
   \caption{\textsf{\textbf{Parameter space and time evolution of retained water ice in planetesimal simulations}. Same as Fig.~\ref{fig:s4}, but for the retained volumetric fraction of water ice (thermochemical criterion $T < T_\mathrm{hydr}$).}}
   \label{fig:s5}
\end{figure*}
\begin{figure*}[tbh]
   \centering
   \includegraphics[width=0.49\textwidth]{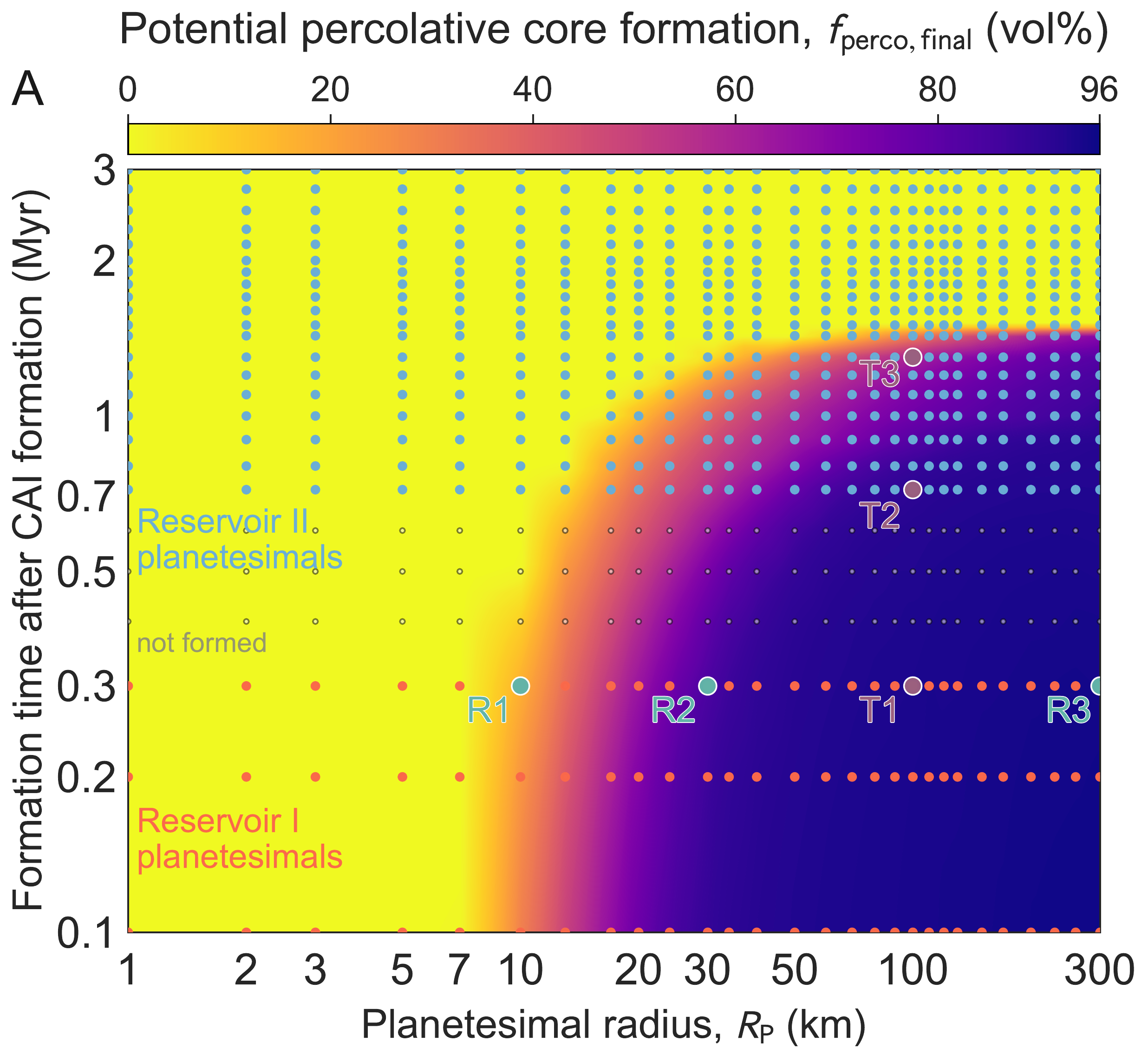}
   \includegraphics[width=0.49\textwidth]{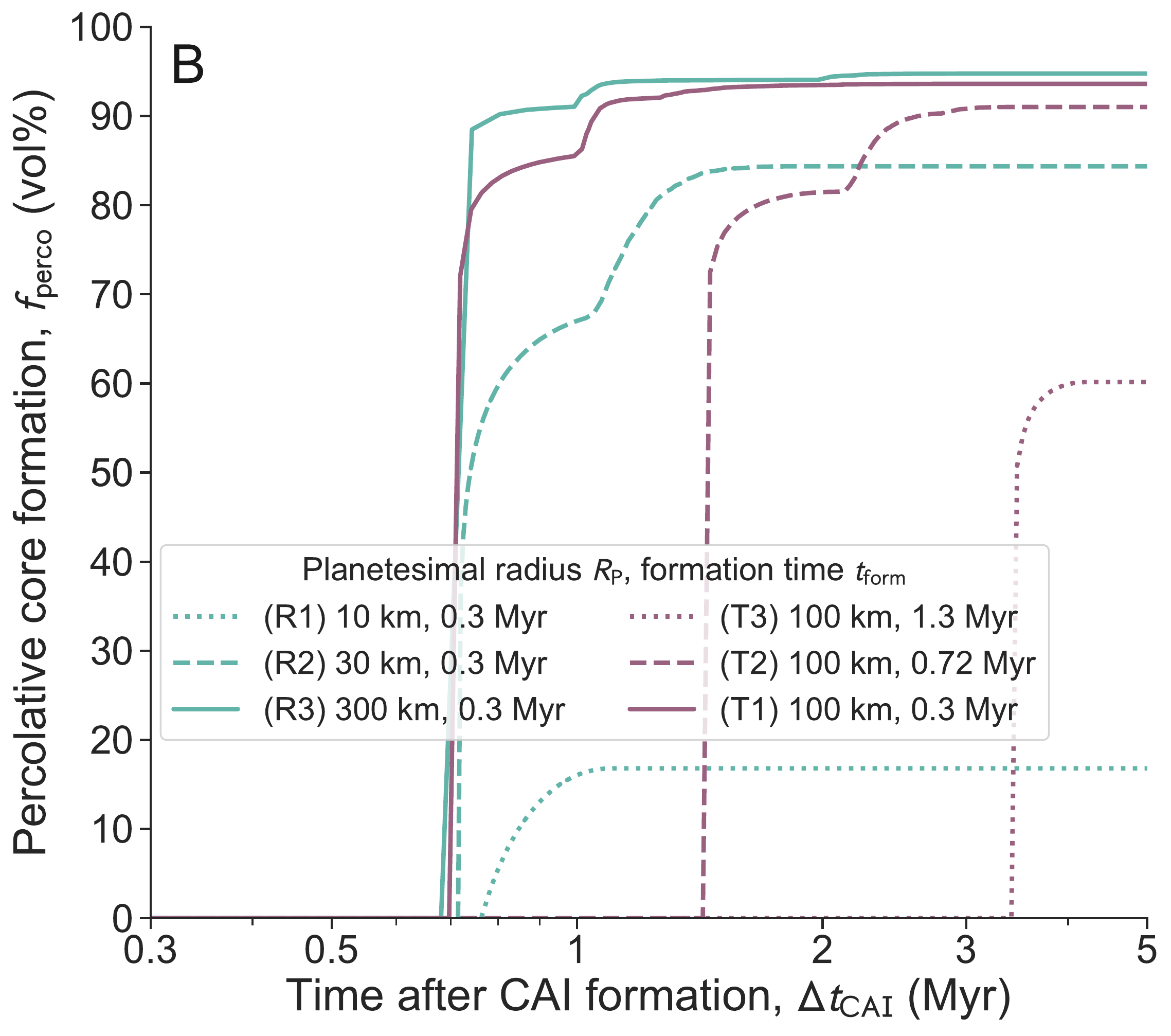}
   \caption{\textsf{\textbf{Parameter space and time evolution of core formation via metal percolation in planetesimal simulations}.  Same as Fig.~\ref{fig:s4}, but for the volumetric fraction eligible for percolative core formation (thermochemical criterion $T \geq T_\mathrm{perc}$).}}
   \label{fig:s6}
\end{figure*}
\begin{figure*}[tbh]
   \centering
   \includegraphics[width=0.49\textwidth]{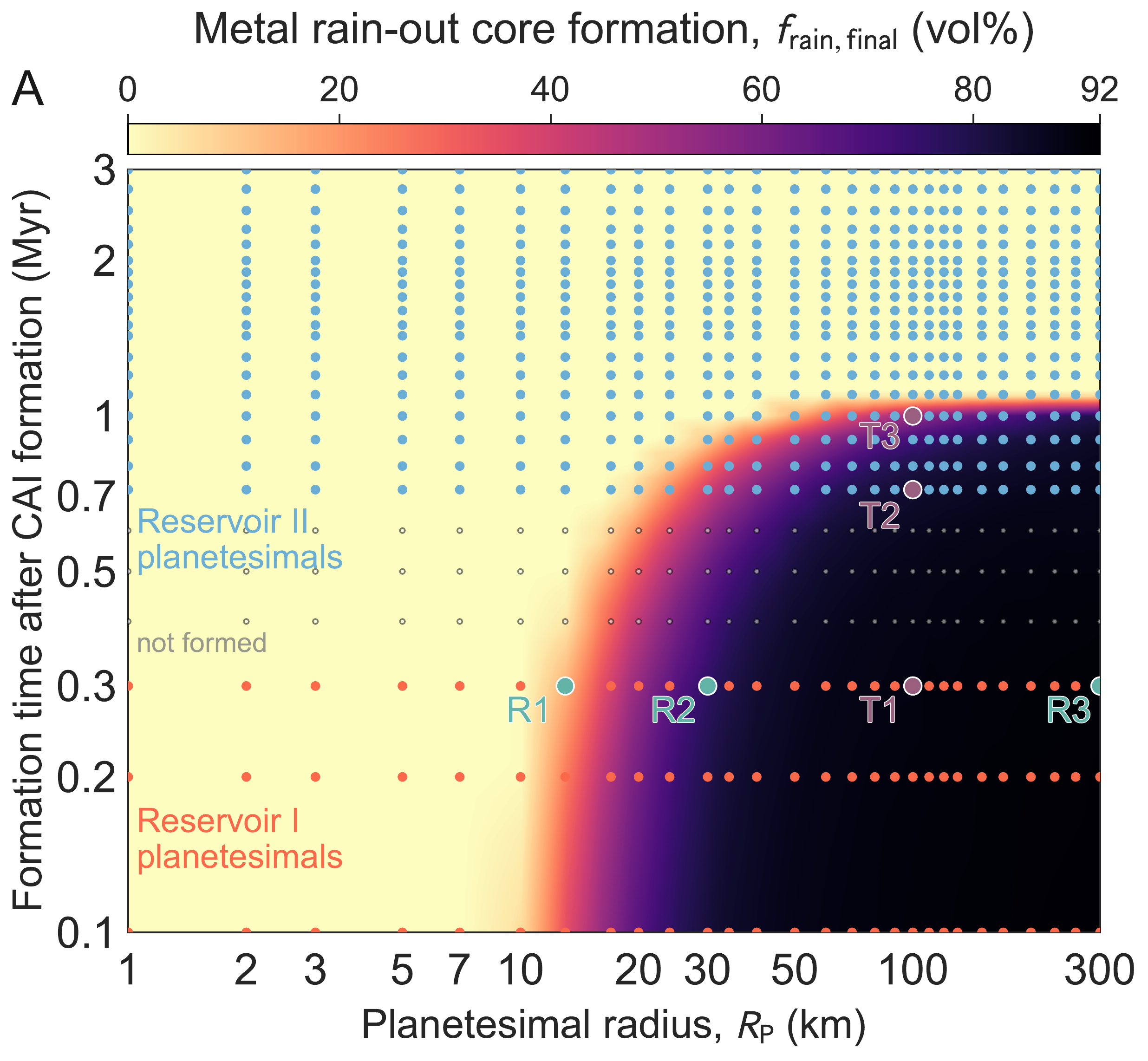}
   \includegraphics[width=0.49\textwidth]{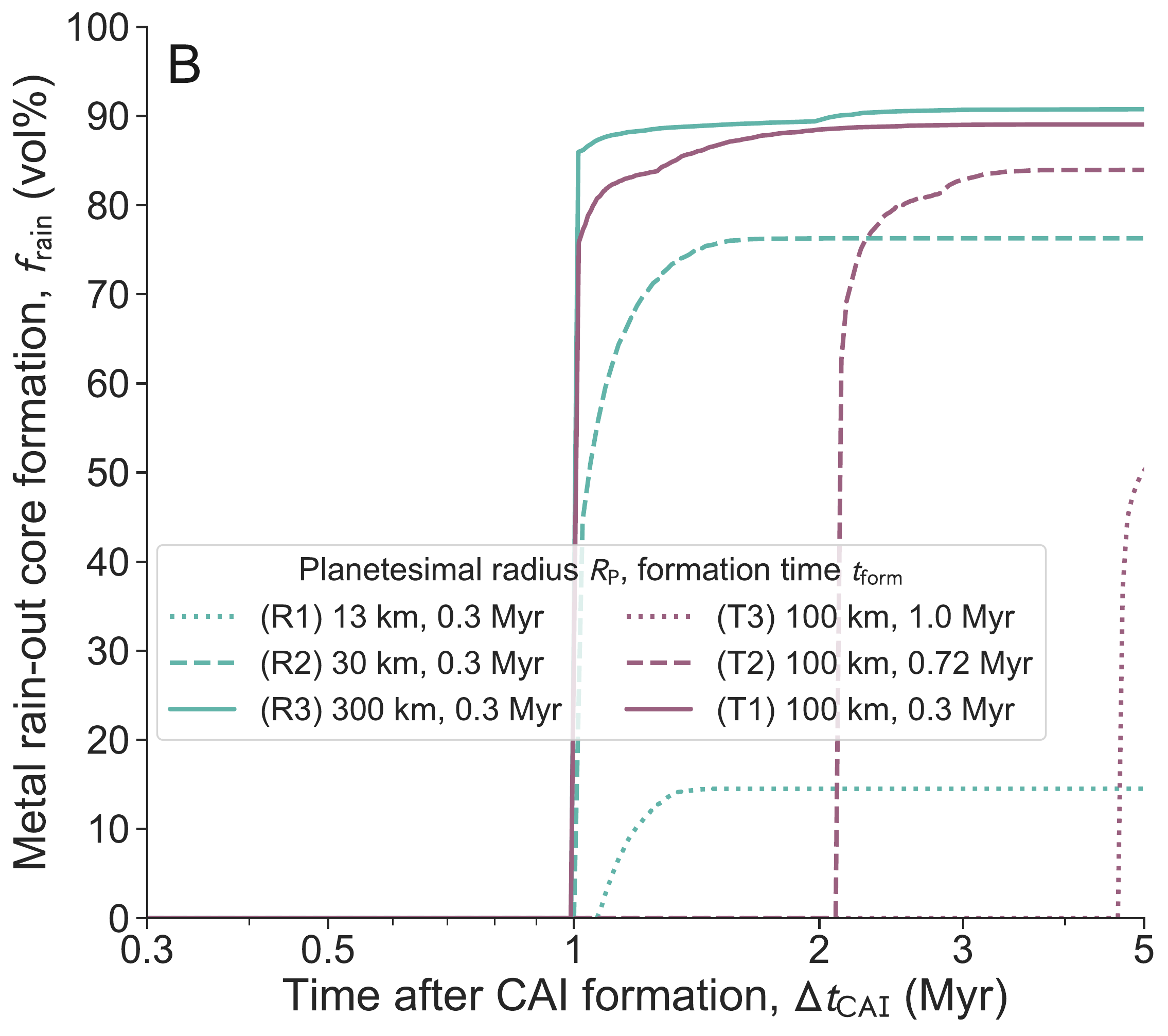}
   \caption{\textsf{\textbf{Parameter space and time evolution of core formation via metal rain-out in planetesimal simulations}. Same as Fig.~\ref{fig:s4}, but for the volumetric fraction eligible for rain-out core formation (thermochemical criterion $\varphi \geq \varphi_\mathrm{rain}$).}}
   \label{fig:s7}
\end{figure*}
\begin{figure*}[tbh]
   \centering
   \includegraphics[width=0.79\textwidth]{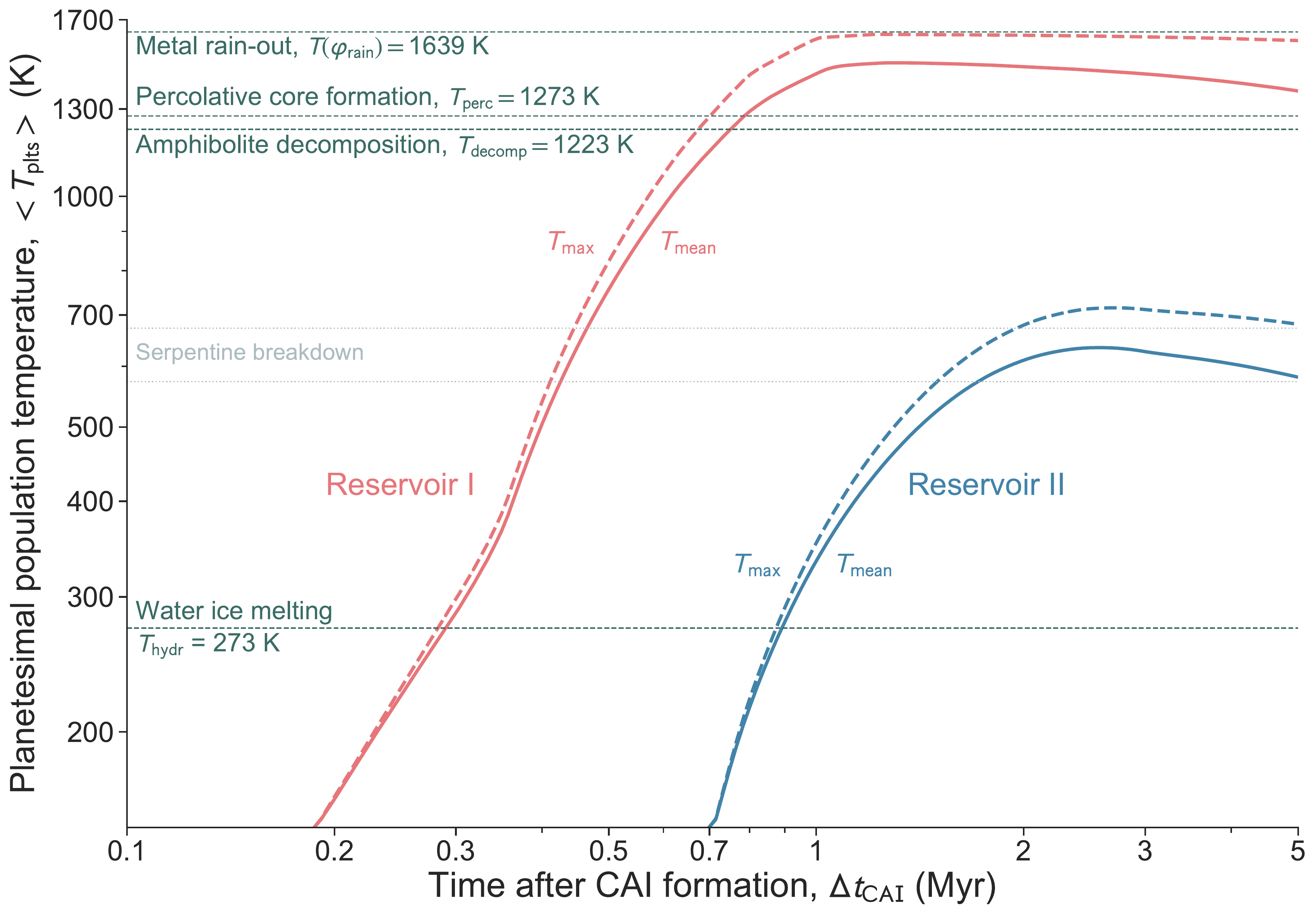}
   \caption{\textsf{\textbf{Temperature evolution of the simulated planetesimal populations.} Shown are the evolution of the volumetric mean (solid lines) and maximum (dashed lines) temperatures over the whole planetesimal population. Reservoir I is indicated in red, Reservoir II in blue. Overplotted are the employed thermochemical criteria for the stability of water ice ($T_\mathrm{hydr}$), hydrous rock ($T_\mathrm{decomp}$), percolative core formation ($T_\mathrm{perc}$), and rain-out core formation ($\varphi_\mathrm{rain}$) with dark green dashed lines. Additionally we indicate the breakdown for more common serpentine phyllosilicate phases (gray dotted lines, labeled 'Serpentine breakdown') in the interval of $T \in$ [573--673] K \citep{2006E&PSL.242...26N,2008EP&S...60..855N}.}}
   \label{fig:s8}
\end{figure*}
Once planetesimals are formed in the disk model, we follow their internal evolution using published methodology \citep{2018Icar..302...27L,2019NatAs...3..307L}. We consider single planetesimals of radius $R_{\mathrm{P}}$ and instantaneous formation time $t_{\mathrm{form}}$. The formation time defines the planetesimal internal heat budget assuming a disk-wide initial homogeneous ratio of ($^{\mathrm{26}}$Al/$^{\mathrm{27}}$Al)$_{\mathrm{CAI}}$ $\equiv 5.25 \times 10^{-5 }$ \citep{2013M&PS...48.1383K} at the time of CAI formation \citep{2012Sci...338..651C}, which defines our time zero.

Our numerical model of the thermochemical evolution of planetesimals builds on the fluid mechanical framework \texttt{I2ELVIS} \citep{2003PEPI..140..293G,2007PEPI..163...83G,gerya2019book}. We numerically solve the fluid dynamic conservation equations using the extended Boussinesq approximation \citep{gerya2019book} in a two-dimensional infinite cylinder geometry on a Cartesian grid, namely the continuity, Stokes, Poisson, and energy conservation equations,
\begin{eqnarray}
\frac{\partial \rho}{\partial t} + \nabla \rho \mathbf{v} &=& 0,\\
\nabla \mathbf{\sigma'} - \nabla P + \rho \mathbf{g} &=& 0,\\
\nabla^2 \Phi &=& 4 \pi G \rho,\\
\rho c_{\mathrm{P}} \left(\frac{\partial T}{\partial t} + v_{\mathrm{i}} \cdot \nabla T \right) &=& - \frac{\partial q_{\mathrm{i}}}{\partial x_{\mathrm{i}}} + H_{\mathrm{r}} + H_{\mathrm{s}} + H_{\mathrm{L}},
\end{eqnarray}
with density $\rho$, flow velocity $\mathbf{v}$, deviatoric stress tensor $\mathbf{\sigma}'$, pressure $P$, directional gravity ${\mathbf{g}}$, gravitational potential $\Phi$, Newton's constant $G$, heat capacity $c_{\mathrm{P}}$, temperature $T$, heat flux $q_{\mathrm{i}} = -k \partial T/\partial x_{\mathrm{i}}$, thermal conductivity $k$, and radiogenic ($H_{\mathrm{r}}$), shear ($H_{\mathrm{s}}$) and latent ($H_{\mathrm{L}}$) heat production terms. Employing a Lagrangian marker-in-cell technique to integrate the energy equation, we minimise numerical diffusion and trace advection of non-diffusive flow properties during material deformation \citep{gerya2019book}. The computational stencil is formulated using a fully staggered-grid finite-differences method, which captures sharp variations of stresses and thermal gradients in models with strongly variable viscosity and thermal conductivity. 

The thermal disk gas boundary conditions of the planetesimals are assumed to be constant in time, $T_{\mathrm{disk}} \equiv 150$ K. Radiogenic heating from $^{\mathrm{26}}$Al decays with time,
\begin{equation}
  H_{\mathrm{r}}(t) = f_{\mathrm{Al}} \cdot \left(\frac{^{\mathrm 26}\mathrm{Al}}{^{\mathrm 27}\mathrm{Al}}\right)_{\mathrm{CAI}} \cdot \frac{E_{^{26}\mathrm{Al}}}{\tau_{^{\mathrm 26}\mathrm{Al}}} \cdot \mathrm{e}^{-\Delta t_{\mathrm{CAI}}/\tau_{^{\mathrm 26}\mathrm{Al}}},
\end{equation}
with the chondritic abundance of aluminum $f_{\mathrm{Al}}$ \citep{2003ApJ...591.1220L}, the decay energy $E_{^{\mathrm 26}\mathrm{Al}} = 3.12$ MeV \citep{2009Icar..204..658C}, the time relative to CAI formation $\Delta t_{\mathrm{CAI}}$, and the $^{\mathrm{26}}$Al mean lifetime $\tau_{^{\mathrm 26}\mathrm{Al}} = 1.03$ Myr \citep{2013M&PS...48.1383K}. The thermal consequences of magma ocean stages beyond the rock disaggregation threshold \citep{2009GGG....10.3010C} are parameterized using a scaled thermal conductivity,
\begin{equation}
k_{\mathrm{eff}} = (q_\mathrm{conv}/0.89)^{3/2} \cdot \alpha_{\mathrm{liq}} g c_{\mathrm{p-Si}} / (\Delta T^2 \rho_{\mathrm{s}} \eta_{\mathrm{num}}),
\end{equation}
with the convective heat flux $q_\mathrm{conv}$, the temperature difference across nodes $\Delta T$, silicate density $\rho_{\mathrm{s}}$, thermal expansivity of molten silicates $\alpha_{\mathrm{liq}}$, silicate heat capacity $c_{\mathrm{p-Si}}$, local gravity $g(x,y)$, and lower cut-off viscosity $\eta_{\mathrm{num}}$. Numerical rock properties and the \texttt{I2ELVIS} treatment of rock melting have been described previously \citep{2016Icar..274..350L}. 

To estimate the compositional evolution of planetesimals that form from an initial mixture of silicate rock, iron metal, and water ice, we follow an updated version of the procedure described in \citep{2019NatAs...3..307L,Monteux2018} and derive the fraction of each planetesimal in our simulation grid that exceed certain thermochemical criteria. We employ four criteria, two related to water ice melting and water-rock reactions, and two related to metal core formation. During radiogenic heat-up of a water-ice-rich planetesimal, the primordial ice melts and may react with the ambient rock to create hydrous silicates and oxidize available iron phases, resulting in loss of hydrogen to space \citep{castillorogez2017,2017GeCoA.211..115S}. While in liquid phase, the aqueous fluid may undergo pore water convection along a down-temperature gradient \citep{1993Sci...259..653G,1999Sci...286.1331Y}, but a fraction of water can be trapped in hydrous silicate phases. Therefore, we derive approximate upper and lower thermal bounds of the thermochemical interior evolution by calculating the fraction of the planetesimal body that exceeds certain threshold temperatures for ice melting/rock hydration at $T \geq T_{\mathrm{hydr}} \equiv$ 273~K, and dehydration and decomposition of the most stable hydrous silicate phases at $T \geq T_{\mathrm{decomp}} \equiv$ 1223~K \citep{Monteux2018}, which we regard as a generous upper limit, because most hydrous phases would be destroyed at much lower temperatures \citep{2014E&PSL.390..128F,2006E&PSL.242...26N,2008EP&S...60..855N}. To estimate the fraction of the body that may undergo metal-silicate segregation, we employ two criteria. If substantial sulfur (S) is present in the parent body, an early, incomplete core may form \citep{2009E&PSL.288...84B} from an interconnected network of metal-sulfide phases in the host body and subsequent gravitational drainage (percolation) toward the center, which we fix to the Fe-FeS eutectic temperature at 1 bar, $T \geq T_{\mathrm{perc}} \equiv$ 1273~K. When temperatures rise further and silicate rocks start to melt they eventually reach the rock disaggregation threshold beyond a melt fraction of $\varphi \gtrsim 0.4$ \citep{2009GGG....10.3010C}, which is equivalent to $\approx$1639~K at 1~bar. Accordingly, at this melt fraction, $\varphi \geq \varphi_\mathrm{rain} \equiv 0.4$, we assume closure of core formation due to efficient iron droplet rain-out from the internal magma ocean \citep{2018Icar..302...27L,2012AREPS..40..113E,2020PhRvF...5k4304P}.

For each timestep in the disk evolution model we then compute the fraction $f_{\mathrm{P,crit}}$ of computational markers in the planetesimal body that breach the aforementioned thresholds,
\begin{equation}
f_{\mathrm{P,crit}}(t) \equiv N_{\mathrm{markers,crit}}/N_{\mathrm{markers,total}},
\end{equation}
with $N_{\mathrm{markers,crit}}$ the number of markers in a planetesimal that exceed the threshold, and $N_{\mathrm{markers,total}}$ the total number of markers in the planetesimal body. Fig.~\ref{fig:3}A shows the peak temperatures for various combinations of planetesimal formation time and radius (as a volumetric mean per body); we show additional parameter ranges in Figs. \ref{fig:s4}--\ref{fig:s7}. In general, larger planetesimal radius and earlier formation time (due to more live $^{26}$Al) lead to higher temperatures in the interior. Peak temperatures are always reached at the center of the planetesimal, which is most insulated from the cold exterior. Typically, temperatures decrease from the inside out. However, because the penetration depth of inward conduction at the planetesimal surface is very limited (see below and \citet{castillorogez2017,2016Icar..274..350L,2019E&PSL.507..154L}), planetesimal interiors above $\approx$ 50 km in size are approximately isothermal in their deeper interiors after an initial heat-up phase. Once high silicate melt fractions beyond the disaggregation threshold are reached, the maximum temperatures are effectively buffered by convective heat transport.

The computational models cover the evolution of single planetesimals. To process the time-dependent evolution of the planetesimals formed in the disk simulation, we follow a procedure \citep{2018Icar..302...27L} to analyze the thermochemical evolution of a planetesimal population using a Monte Carlo approach that parameterizes the birth planetesimal population from the streaming instability mechanism, as indicated by fluid dynamical models \citep{2015SciA....115109J,2017ApJ...847L..12S,2019ApJ...883..192A,2019NatAs...3..808N}. For each timestep in the planetesimal formation model (see above) we randomly generate a family of newly formed planetesimals that follow a radius power law
\begin{equation}
\frac{\partial N_{\mathrm{P}}}{\partial R_{\mathrm{P}}} \propto R_{\mathrm{P}}^{-q},
\label{eq:sfd}
\end{equation}
with the number of bodies in a specific radius bin $N_{\mathrm{P}}$, and power law index $q = 2.8$ \citep{2015SciA....115109J,2017ApJ...847L..12S}, from which we generate integer radii $R_{\mathrm{min}} \leq R_{\mathrm{P}} \leq R_{\mathrm{max}}$ according to
\begin{equation}
R_{\mathrm{P}} = ||R_{\mathrm{min}} (1-x_{\mathrm{rand}})^{-1/(q-1)}||,
\end{equation}
with pseudo-random number $x_{\mathrm{rand}} \in$ [0,1], approximating the total newly generated planetesimal mass per timestep and reservoir using
\begin{equation}
\frac{\partial M_{\mathrm{P,res}}}{\partial t} = \int_{m_{\mathrm{min}}}^{m_{\mathrm{max}}} \mathrm{d} m_{\mathrm{P,res}}(R_{\mathrm{P}},t),
\end{equation}
where the planetesimal mass $m_{\mathrm{P}}$ and the minimum and maximum bounds, $m_{\mathrm{min}}$ and $m_{\mathrm{max}}$, are related to the planetesimal radius by a fixed initial mean density of $\rho_{\mathrm{P}} = 2750$ kg m$^{-3}$. We choose the minimum planetesimal radius in our parameter space to be $R_{\mathrm{min}}$ = 1 km, and set the upper bound of the birth planetesimal population to $R_{\mathrm{max}}$ = 300 km, consistent with a tapered power law toward larger planetesimal radii \citep{2015SciA....115109J,2019ApJ...883..192A} and planetesimal mean birth sizes on the order of $\approx$100 km \citep{2009Icar..204..558M,2017Sci...357.1026D}.

To compute the compositional evolution of the whole planetesimal population per timestep and thermochemical criterion, we bin the planetesimal mass formed in the disk model into distinct radius bins according to the primordial planetesimal size-frequency distribution (Eq. \ref{eq:sfd}). We then calculate the fraction of the planetesimal population per reservoir in a given radius and time bin that fulfills a specific thermochemical criterion as
\begin{eqnarray}
f_{\mathrm{pop,crit}}(t) & = & \frac{1}{N_{\mathrm{P,tot}}(t)} 
\\ & & \cdot \int_{R_{\mathrm{min}}}^{R_{\mathrm{max}}} N_{\mathrm{P,bin}} \cdot \ f_{\mathrm{bin,crit}} \cdot  \mathrm{d}V_{\mathrm{bin}}(t), \nonumber
\end{eqnarray}
with $N_{\mathrm{P,tot}}(R_{\mathrm{P}},t)$ the total number of planetesimals per reservoir, $N_{\mathrm{P,bin}}$ the number of planetesimals per bin, and $f_{\mathrm{bin,crit}}(R_{\mathrm{P}},t) \cdot \mathrm{d}V_{\mathrm{bin}}(t)$ the volumetric fraction, linearly interpolated from the single planetesimal simulation grid described above. The volume change per timestep is normalized to the total planetesimal volume per reservoir and time after 5 Myr in the disk simulation. The volumetric fraction increases when more planetesimals form or when a planetesimal sub-volume starts to satisfy a specific criterion (e.g., hot enough for percolative core formation), and decreases when a planetesimal sub-volume does not satisfy a given criterion anymore (e.g., too hot for water ice stability). In Fig. \ref{fig:s8} we display the combined temperature evolution of the planetesimal populations formed in the disk model in Reservoir I and Reservoir II together with the thermochemical criteria. The effects of minimum and maximum radii and choice of power law have been discussed elsewhere \citep{2018Icar..302...27L}.

\subsection*{Exterior heating effects on planetesimals due to disk evolution}
\label{sec:1d_planetesimal_models}
\begin{figure*}[tbh]
   \centering
   \includegraphics[width=0.79\textwidth]{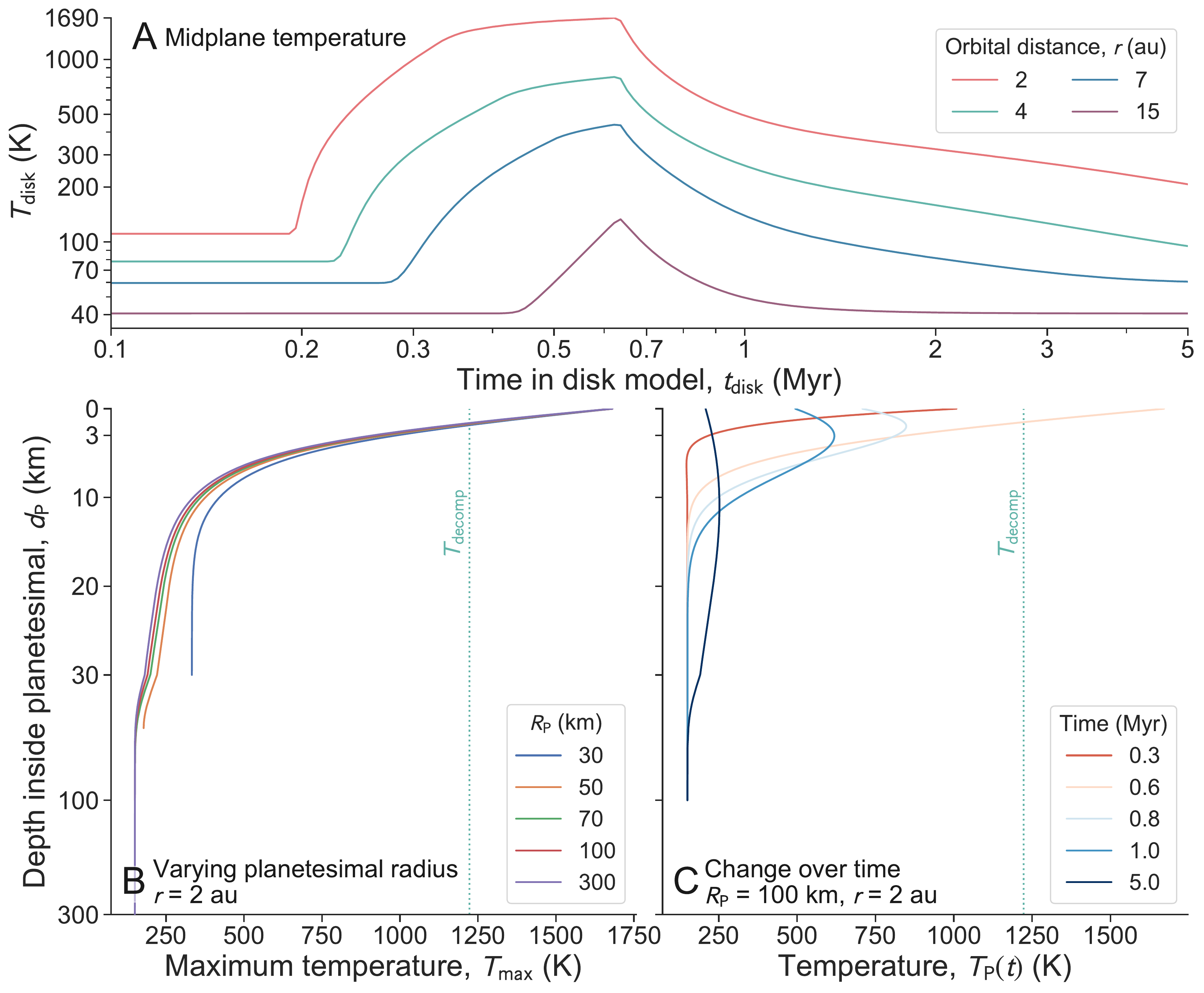}
   \caption{\textsf{\textbf{Effects of external disk temperature on the thermal profile in planetesimals.} (\textbf{A}) Midplane gas temperature for 2, 4, 7, and 15 au in the disk simulation (lines). (\textbf{B}) Resulting peak temperature profiles in planetesimals between 30--300 km in radius at 2 au. The dashed green line shows the temperature criterion for hydrous rock decomposition ($T_\mathrm{decomp}$). (\textbf{C}) Time evolution for a 100 km radius planetesimal at 2 au. Different line colors indicate different times during the disk lifetime.}}
   \label{fig:s10}
\end{figure*}
Our planetesimal models employ a constant temperature boundary condition, whereas in the disk simulations the local gas temperature can vary over time (Fig. \ref{fig:s2}). In general, the exterior disk temperature has minor effects on planetesimal evolution \citep{2014M&PS...49.1083G}. In this section we seek to determine whether it is negligible in our simulations. To estimate the effect of the time-dependent disk temperature on the thermal evolution of planetesimals, we developed a thermal diffusion code that solves the radial 1-D heat diffusion equation \citep{2006M&PS...41...95H,gerya2019book},
\begin{eqnarray}
\frac{\partial T}{\partial t} = \frac{k}{\rho c_\mathrm{P}} \frac{1}{r^2}\frac{\partial}{\partial r}\left(r^2\frac{\partial T}{\partial r}\right),
\end{eqnarray}
where $T$ is the temperature, $t$ is the time, $k$ is the thermal conductivity, $\rho$ is the density, $c_\mathrm{P}$ is the heat capacity and $r$ is the radial distance from the center. We solve the heat diffusion equation using the implicit finite difference method \citep{gerya2019book}. For consistency with the planetesimal calculations above, we assume that thermal conductivity, density and heat capacity are all temperature-independent and use the same values as employed in the more complex planetesimal models. As here we are only interested in the effect of the disk temperature on planetesimal evolution, we do not consider radiogenic heating. The initial temperature of the planetesimal interior is the same as in the previously described models. 

To simulate the influence of the disk temperature on the planetesimal interior we prescribe the time-dependent disk temperature at 2 au as surface boundary condition, the orbital location with the highest temperatures where planetesimals form in the disk model (Fig. \ref{fig:s10}A). We model a profile across the entire planetesimal diameter over time, covering the various evolution stages of the disk. To check numerical convergence of this simplified code, we performed calculations with different resolutions ranging from 1000 to 8000 grid points and tested timestep lengths ranging from 25 days to 1 yr. The results (Fig. \ref{fig:s10}) show the code has converged and demonstrate that disk heating of the planetesimals is limited to the near surface layers, thus only plays a volumetric role for small planetesimals, while the interiors of larger planetesimals remain mostly unaffected. Maximum temperatures achieved at various depths inside differently sized planetesimals are plotted in Fig. \ref{fig:s10}B and the time-evolution of temperature for the $R_\mathrm{P}$ = 100 km case is displayed in Fig. \ref{fig:s10}C. These are the maximum possible effects in the disk model. Internal temperatures of bodies formed at more distant orbits (Fig. \ref{fig:s10}A) are even less affected.

\subsubsection*{Planetary migration}
\label{sec:migration}

To estimate the migration timescales for planetary embryos in Reservoir I and II (Fig. \ref{fig:1}), we compute the torque from the disk on an embryo at each timestep and use this to evolve the total angular momentum of an embryo of fixed mass with a first-order integrator. For type I migration (see e.g. \citet{Baruteau} for a review) the torque scales as
\begin{equation} \label{eq:gam0}
\Gamma_0 = \frac{q_{\mathrm{P}}^2}{h^2}  \Sigma_{\mathrm g}  r^4  \Omega_{\mathrm p}^2,
\end{equation}
with the embryo-to-star mass ratio $q_{\mathrm{P}} = M_{\mathrm{embryo}}/M_{\star}$, disk aspect ratio $h$, and Keplerian frequency $\Omega_{\mathrm p}$. The exact magnitude and direction of the torque can be written as a sum of the so-called Lindblad or `wave' torque -- generated by waves driven by the planet at resonant locations in the disk  --  and the corotation torque -- generated by gas that is on average corotating with the planet. The wave torque \citep{Paardekooper2010} can be written as 
\begin{equation}
    \gamma {\Gamma_{\mathrm{wave}}}/{\Gamma_0} =  - 2.5 - 1.7 B + 0.1 A,
\end{equation}
where $\gamma$ is the ratio of specific heats, $B$ is the negative power law exponent of the local disk temperature profile and $A$ is the negative power law exponent of the local disk density profile.

The calculation of the corotation torque requires determining how both diffusion within the gas and planetary mass affect the material in this region. At a given viscosity, the torque peaks in magnitude as the non-linear `horseshoe drag'. With too little viscosity to replenish the angular momentum within this region,  the horseshoe drag can become saturated, leaving only the Lindblad torque. For sufficiently high viscosity, the torque is forced into a linear regime which is reduced in magnitude. We therefore account for local disk conditions when computing the overall magnitude and direction of the torque. The linear and non-linear versions of the barotropic component of the torque \citep{Paardekooper2010} are given by 
\begin{eqnarray}
    \gamma {\Gamma_{\mathrm{lin,baro}}}/{\Gamma_0} &=&  0.7  (3/2 - A), \\
    \gamma {\Gamma_{\mathrm{hs,baro}}}/{\Gamma_0} &=&  1.1  (3/2 - A)
\end{eqnarray}
respectively, while the entropy-related components of the linear corotation torque and horseshoe drag are given by
\begin{eqnarray}
     \gamma {\Gamma_{\mathrm{lin,ent}}}/{\Gamma_0} &=& (2.2 - 1.4/\gamma) \epsilon_\mathrm{e},\\
     \gamma {\Gamma_{\mathrm{hs,ent}}}/{\Gamma_0} &=& 7.9 \epsilon_\mathrm{e}/\gamma,
\end{eqnarray}
respectively, with $\epsilon_\mathrm{e} = B - (\gamma - 1)A$ being the negative of the local power law exponent of the specific entropy profile.

To choose $\gamma$, we utilise the fact that the temperature of the disk is set only by radiative and viscous heating, and radiative cooling. We expect the radiative processes to be much faster than the dynamical timescale, so consider the disk to be locally isothermal and use $\gamma = 1$. This assumption implicitly adds infinite thermal diffusion to the gas, driving the entropy-related part of the corotation torque into the linear regime. For transitioning between the unsaturated and saturated states of the torque, we simply switch various components of the corotation/horseshoe drag on and off as necessary -- we consider the overall value of the torque $\Gamma_{\mathrm{tot}}$ to be defined by one of three regimes, largely governed by the behaviour of the barotropic component:
\begin{itemize}
    \item `Unsaturated', in which the total torque is the sum of the wave torque, the linear entropy-related torque and the non-linear barotropic horseshoe drag ($\Gamma_{\mathrm{tot}} = \Gamma_{\mathrm{wave}} + \Gamma_{\mathrm{lin,ent}}  + \Gamma_{\mathrm{hs,baro}}$).
    \item `Saturated', in which the corotation region is assumed to be totally depleted of angular momentum such that the planet feels no torque from this region and the only remaining component is the wave torque ($ \Gamma_{\mathrm{tot}} = \Gamma_{\mathrm{wave}}$).
    \item `Linear', in which the total torque is the sum of the wave torque and the linear forms of both the entropy-related and barotropic corotation torque ($ \Gamma_{\mathrm{tot}} = \Gamma_{\mathrm{wave}} +  \Gamma_{\mathrm{lin,ent}} + \Gamma_{\mathrm{lin,baro}}$).
\end{itemize}
We must then choose one of these three regimes at each timestep. The maximal, non-linear, unsaturated horseshoe drag is achieved when the disk viscosity parameter $\alpha_{\mathrm{v}}$ satisfies \citep{BaruteauLin}
\begin{equation}\label{eq:regime}
0.16 \frac{q_{\mathrm{p}}^{3/2}}{h^{7/2}} < \alpha_{\mathrm{v}} < 0.16 \frac{q_{\mathrm{p}}^{3/2}}{h^{9/2}}.
\end{equation}
Values of $\alpha_{\mathrm{v}}$ lower than the left-hand side of this criterion will cause the torque to become saturated, while values higher than the right-hand side will push the torque into its linear form. At each timestep we select a regime for the torque based on which of the inequalities in Eq. \ref{eq:regime} is satisfied. We compute the total torque based on the regime and the value of $\Gamma_0$, using Eq. \ref{eq:gam0}  and the disk properties taken directly from the disk model. We linearly interpolate between different snapshots of this disk model to obtain values for disk properties on timescales relevant for migration. 
\begin{figure*}[tb]
   \centering
   \includegraphics[width=0.95\textwidth]{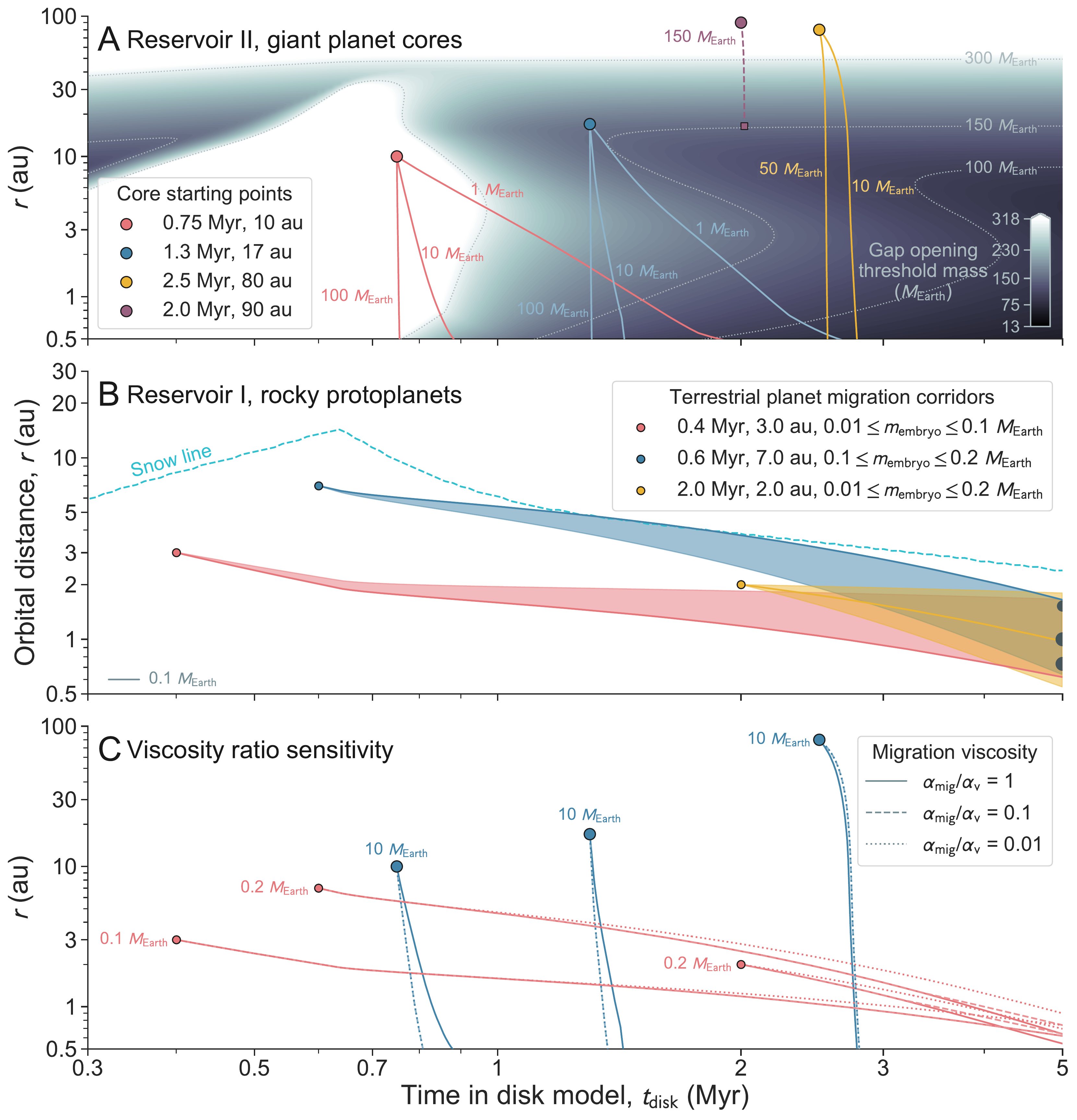}
   \caption{\textsf{\textbf{Type I and type II orbital migration in the disk simulation.} (\textbf{A}) Migration of massive giant planet cores from starting locations and times (red, blue, yellow, and purple circles) located in Reservoir II. Lines show the migration path for a given mass (1, 10, 50, 100, or 150 Earth masses). The colored background indicates the mass threshold for a giant planet core to open a gap in the disk gas and transition from the type I to the type II migration regime (like the purple 150 $M_\mathrm{Earth}$ core). (\textbf{B}) Type I migration for various terrestrial planet embryo masses and starting locations and times (red, blue, and yellow circles) in Reservoir I. Lines and shades indicate the migration paths for a given starting location and protoplanet mass between 0.01--0.1 $M_\mathrm{Earth}$ (red), 0.1--0.2 $M_\mathrm{Earth}$ (blue), or 0.01--0.2 $M_\mathrm{Earth}$ (yellow). Planets that grow more massive during the disk lifetime would migrate inward of 0.5 au. (\textbf{C}) Disk viscosity sensitivity of the shown simulation tracks. For each starting location and time in Reservoir I (red circles) or II (blue circles), the disk viscosity that drives type I migration ($\alpha_\mathrm{mig}$) is varied between $\alpha_\mathrm{v}$ (solid lines) and $\alpha_\mathrm{t}$ (dotted lines).}}
   \label{fig:s9}
\end{figure*}
As a test of the validity of this model, we compared the estimated migration timescales with those in two hydrodynamical simulations of a migrating Earth-mass planet using the two-dimensional \texttt{FARGO-3D} code \citep{Llambay2016}. In both tests the planet started at 1.8 au with $\alpha_{\mathrm{v}} = 1\times 10^{-4}$, a surface density profile $\Sigma = (1000 \mathrm{g/cm}^3)r^{-1}$ and $h \propto r^{1/4}$. We created these initial conditions in both model setups. In the first test we used $h = 0.073$ at 1 au, placing our model in the `linear' regime. The difference between the migration rate in our model and the \texttt{FARGO-3D} simulations was around 7\% over the 2000 yr simulation time. In the second test we used $h = 0.035$ (the `unsaturated' regime) and found that the migration rate in the simulation was 32\% faster than in our model. The parameters of the latter test would -- according to the model in \citet{Paardekooper2011} -- partially saturate the corotation torque in our disc. Because the corotation torque promotes outward migration in this disc, the partial saturation would result in faster migration in the hydrodynamical models as observed, so any difference between our model and equivalent hydrodynamical simulations is due to this approximate treatment of saturation. We conclude that a more sophisticated approach would not alter the conclusions of this paper.

In addition to the type I migration timescales we investigate whether a migrating protoplanet opens a gap in the disk, which may slow down its migration. To determine if and when this happens, we use the gap-opening criterion $P < 1$ \citep{Crida}, where
\begin{equation}
    P = \frac{3^{4/3}}{4}\frac{h}{q_{\mathrm{P}}^{1/3}} +\frac{50 \alpha_{\mathrm{v}} h^2}{q_{\mathrm{P}}}.
\end{equation}
We consider any embryo which satisfies $P < 1$ to be gap opening and calculate the threshold masses for each orbital radius and timestep to estimate whether a protoplanet in Reservoir II may be halted in its migration within the context of our disk model.

Figs. \ref{fig:1} and \ref{fig:s9} show migration tracks from this model. Fig. \ref{fig:s9}A shows the migration tracks of potentially early-formed giant planet cores in Reservoir II. In all but the 150 Earth mass ($M_\mathrm{Earth}$) case, type I migration is too rapid to open up a gap and transition to the (slower) type II regime. In more comprehensive simulation setups, migration and growth show a complex interplay \citep{2019A&A...623A..88B,2019A&A...624A.109B,2019A&A...622A.202J}, but our models illustrate that early-formed giant planet cores typically migrate rapidly toward the star. Fig. \ref{fig:s9}B illustrates that also smaller-mass planets, such as the embryos that accrete to form the inner terrestrial planets, are limited in their maximum mass accretion during the disk phase. If they grow too fast and become too massive, they migrate rapidly inward. Later formation beyond the snow line (to reach the present-day orbits) would result in overly water-rich planets \citep{2019NatAs...3..307L,2019A&A...624A.109B}, inconsistent with the present-day volatile inventory of the Solar System terrestrial planets \citep{PeslierISSI2018}. Fig. \ref{fig:s9}C displays the small sensitivity of the migration efficiency on the employed midplane viscosity. Whether protoplanets in the type I regime are migrating with $\alpha_\mathrm{v}$ or $\alpha_\mathrm{t}$ (see above) has little effect on the simulations.

Our model of migration is simplified by necessity and there are effects which cannot be accounted for in such a model that may affect the dynamical evolution of migrating protoplanets \citep{2019SAAS...45..151K}, such as the heating torque \citep{2015Natur.520...63B,2019A&A...624A.109B}, variations in angular momentum transport in wind- or turbulence-driven disks \citep{2020A&A...633A...4K}, the interplay between growth and migration \citep{2019A&A...622A.202J}, and dynamical interactions between multiple migrating embryos \citep{2014MNRAS.445..749H,2018MNRAS.474.3998H}.

\subsection*{Planetary accretion mode}
\label{sec:accretion}

To determine the timescales for the onset of planetary accretion from an initially gravitationally collapsed population of planetesimals (Fig. \ref{fig:2}), we employ theoretical prescriptions for planetesimal-planetesimal collisional growth \citep{2004ApJ...616..567I} and pebble accretion \citep{2018A&A...615A.138L,2018A&A...615A.178O}.

In the scenario of planetary growth via ballistic accretion of planetesimals \citep{1972epcf.book.....S,2012AREPS..40..251M}, the growth rate of the planetary embryos is determined by the velocity dispersion in the planetesimal population, which is equivalent to the eccentricity distribution. Eccentricity excitation is damped by gas drag, but less so for lower-mass planetesimals, which are excited by larger bodies. We use a previously-derived prescription for the growth timescale (i.e., the characteristic time for doubling the initial mass of a planetary object) in this scenario \citep{2002ApJ...581..666K}. In this context, the growth time scales as
\begin{eqnarray}
\tau_{c, \mathrm{acc}} & \simeq & 1.2 \times 10^{5}\left(\frac{\Sigma_{\mathrm d}}{10 \, \mathrm{g} \, \mathrm{cm}^{-2}}\right)^{-1}\left(\frac{r}{1 \, \mathrm{au}}\right)^{1 / 2} \nonumber\\
&& \times \left(\frac{M_\mathrm{embryo}}{M_\mathrm{Earth}}\right)^{1 / 3} \times  \left(\frac{M_{\star}}{M_{\odot}}\right)^{-1 / 6} \nonumber\\
&& \times \bigg[\left(\frac{\Sigma_{\mathrm g}}{2.4 \times 10^{3} \, \mathrm{g} \, \mathrm{cm}^{-2}}\right)^{-1 / 5} \nonumber\\
&& \times\left(\frac{r}{1 \, \mathrm{au}}\right)^{1 / 20} \times \left(\frac{m_\mathrm{P}}{10^{18} \, \mathrm{g}}\right)^{1 / 15}\bigg]^{2} \mathrm{yr},
\label{eq:plts_growth}
\end{eqnarray}
with the solid (dust) surface density $\Sigma_{\mathrm{d}}$, semi-major axis $r$, mass of the accreting planetary embryo $M_{\mathrm{embryo}}$, mass of the central star $M_{\star}$, mass of the Sun $M_{\odot}$, gas surface density $\Sigma_{\mathrm{g}}$, and mass of the accreting low-mass planetesimals $m_{\mathrm{P}}$. We determine the timescale for the onset of accretion from an initial planetesimal population produced from the streaming instability, which may be the bottleneck for the accretion of larger Mars-sized protoplanets within the disk lifetime \citep{2016A&A...586A..66V,2019A&A...624A.114L}. Following estimates of the initial mass function of planetesimals (IMF) \citep{2015SciA....115109J,2019ApJ...883..192A}, we choose $M_{\mathrm{embryo}} = 3.11 \times 10^{20}$~kg and $m_{\mathrm{P}} = 1.44 \times 10^{18}$~kg. Using mean densities of $\rho_{\mathrm{P}} = 2750$~kg~m$^{-3}$, this translates to radii of $R_{\mathrm{embryo}} = 300$~km and $R_{\mathrm{P}} = 50$~km, respectively. All other parameters for the growth timescale are calculated from the time-dependent evolution of the disk and dust coagulation model, as described above. Planetesimal growth becomes faster when more of the total mass is distributed in smaller objects, as they are more easily dynamically excited. In the chosen IMF, the majority of the planetesimal mass is in large objects, which we approximate by $R_{\mathrm{P}} = 50$ km. However, because of the $1/15$-exponent in Eq. \ref{eq:plts_growth}, the simulation results are only weakly affected by this choice.  
\begin{figure*}[tbh]
   \centering
   \includegraphics[width=0.89\textwidth]{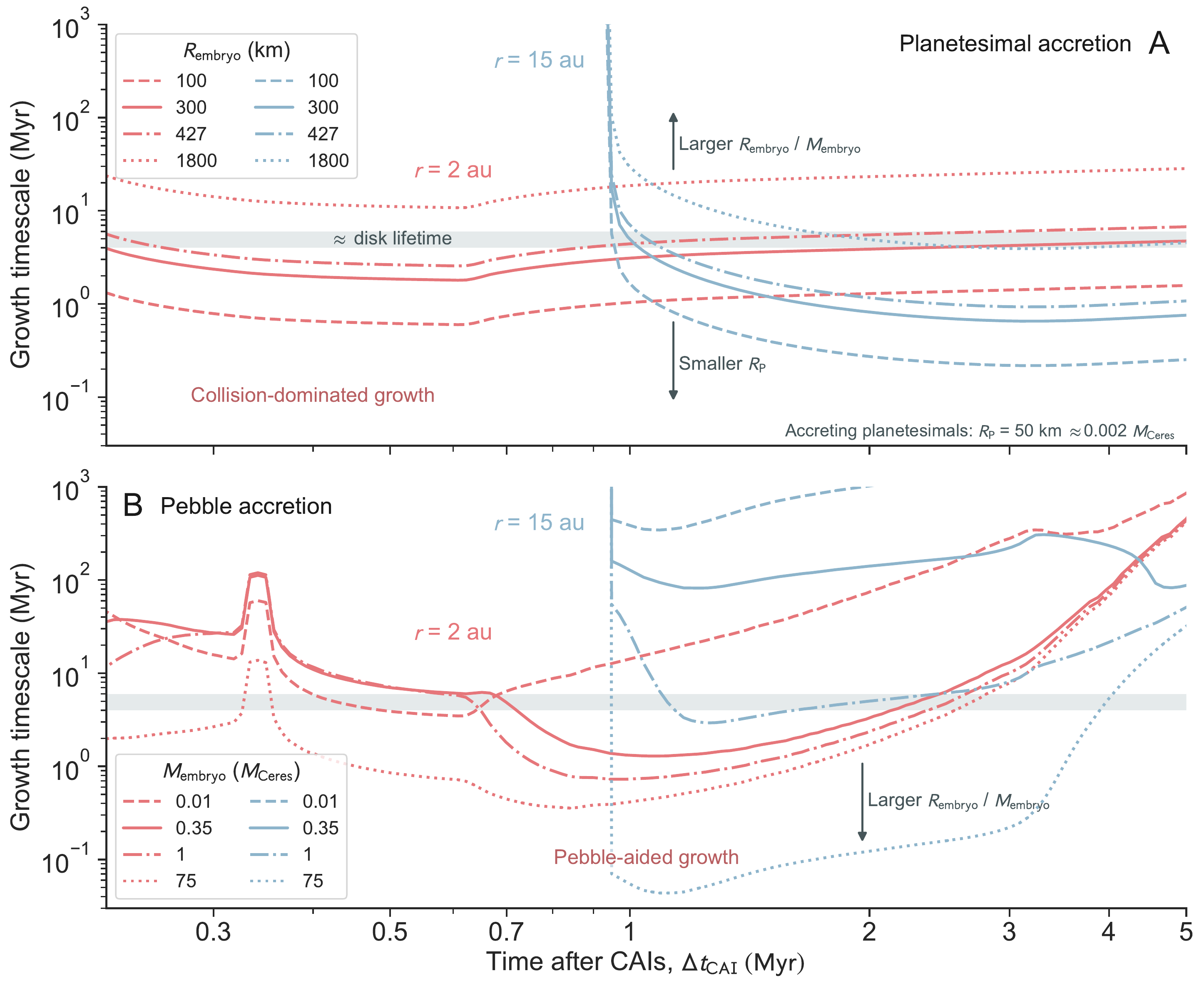}
   \caption{\textsf{\textbf{Growth timescale sensitivity for pebble and planetesimal accretion in Reservoir I and Reservoir II.} The embryo radii and masses (legends) indicated in \textbf{A} and \textbf{B} are equivalent for constant planet density. (\textbf{A}) Growth timescale in the planetesimal (collision) accretion scenario for different embryo radii (dashed, solid, dash-dotted, and dotted lines). Red and blue indicate growth at 2 or 15 au, respectively. The gray horizontal band (labeled '$\approx$ disk lifetime') indicates the approximate lifetime of the circum-solar disk. The up- and downward-pointing gray arrows indicate the trends for the growth timescale: increase for increasing embryo radius $R_\mathrm{embryo}$, decrease for smaller planetesimals ($R_\mathrm{P}$) that accrete onto the embryo. (\textbf{B}) Same as in panel \textbf{A}, but for pebble accretion. The growth timescale decreases for increasing protoplanet mass, but is highly sensitive to secular changes of the pebble flux in the disk simulation.}}
   \label{fig:s11}
\end{figure*}
The timescale for the growth from pebble accretion \citep{2010A&A...520A..43O,2012A&A...544A..32L} is estimated as \citep{2018A&A...615A.138L,2018A&A...615A.178O}
\begin{equation}
    \tau_{\mathrm{c, PA}} = M_{\mathrm{embryo}} / \left(\epsilon_{\mathrm{PA}} \dot{M}_{\mathrm{pebble}}\right),
\end{equation}
with the pebble flux $\dot{M}_{\mathrm{pebbles}}$ and the pebble accretion efficiency 
\begin{equation}
\epsilon_{\mathrm{PA}}=f_{\mathrm{set}} \epsilon_{\mathrm{set}}+\left(1-f_{\mathrm{set}}\right) \epsilon_{\mathrm{bal}},
\end{equation}
which approximates the effects of vertical turbulence on the dust particles in the ballistic and settling regimes of pebble accretion, depending on the settling fraction of individual pebbles $f_{\mathrm{set}}$ \citep{2018A&A...615A.178O}. The efficiencies in the settling regime are given by
\begin{eqnarray}
\epsilon_{\mathrm {set }}&=&\left(\epsilon_{\mathrm{set}, 2 {\mathrm{D}}}^{-2}+\epsilon_{\mathrm{set},3 \mathrm{D}}^{-2}\right)^{-1 / 2},\\
\epsilon_{\mathrm{set},2 \mathrm{D}}&=&\frac{A_{2}}{\eta_\mathrm{disk}} \sqrt{\frac{q_{p}}{\tau_{s}} \frac{\Delta v}{v_{K}}} f_{\mathrm{set}},\\
\epsilon_{\mathrm{set},3 \mathrm{D}}&=&A_{3} \frac{q_{p}}{\eta_\mathrm{disk} h_{\mathrm{eff}}} f_{\mathrm{set}}^{2},
\end{eqnarray}
with the embryo-to-star mass ratio $q_{\mathrm{P}}$, the disk radial pressure gradient $\eta_\mathrm{disk}$, dimensionless stopping time $\tau_{\mathrm{s}}$, relative velocity between pebble and embryo $\Delta v$, Keplerian velocity 
\begin{equation}
v_{\mathrm{K}} = \sqrt{G\left(M_{\star}+M_{\mathrm{embryo}}\right) / a},
\end{equation}
effective pebble disk aspect ratio $h_{\mathrm{eff}}$, and numerical fit constants $A_{\mathrm{2}} = 0.32$ and $A_{\mathrm{3}} = 0.39$ \citep{2018A&A...615A.178O}. The pebble accretion efficiencies in the ballistic regime are given by
\begin{eqnarray}
\epsilon_{\mathrm{bal}, 2 \mathrm{D}}&=&\frac{R_{\mathrm{embryo}}}{2 \pi \tau_{\mathrm{s}} \eta_\mathrm{disk} a} \left(1-f_{\mathrm{set}}\right) \nonumber \\
&& \times \sqrt{\frac{2 q_{\mathrm{p}} a}{R_{\mathrm{embryo}}}+\left(\frac{\Delta v}{v_{\mathrm{K}}}\right)^{2}},\\
\epsilon_{\mathrm{bal}, 3 \mathrm{D}}&=&\frac{1}{4 \sqrt{2 \pi} \eta_\mathrm{disk} \tau_{\mathrm{s}} h_{\mathrm{P}}} \left(1-f_{\mathrm{set}}^{2}\right) \nonumber \\
&& \times \left(2 q_{\mathrm{p}} \frac{v_{\mathrm{K}}}{\Delta v} \frac{R_{\mathrm{embryo}}}{a}+\frac{R_{\mathrm{embryo}}^{2}}{a^{2}} \frac{\Delta v}{v_{\mathrm{K}}}\right),
\end{eqnarray}
with the pebble disk aspect ratio $h_{\mathrm{P}}$ \citep{2018A&A...615A.138L}. The direct dependence on the disk aspect ratio in the above formulation explains why pebble accretion in Fig.~\ref{fig:2} is less efficient in the outer compared to the inner disk \citep{2018A&A...615A.178O}, which we discuss below. The pebble flux $\dot{M}_{\mathrm{pebble}}$ is derived from the disk and dust coagulation model \citep{DD18}. We assume that the growing embryos have no inclination. The pebble accretion efficiency $\epsilon_{\mathrm{PA}}$ decreases with increasing inclination, so the pebble accretion timescale would be longer for an accreting embryo with non-zero inclination. As a compromise between the midplane and global viscosities ($\alpha_{\mathrm{t}}$ and $\alpha_{\mathrm{v}}$, respectively) in the disk model, we choose a vertical diffusion of $\alpha_{\mathrm z} = 1 \times 10^{-4}$ \citep[cf.][]{2018A&A...615A.178O}.

Fig. \ref{fig:s11} shows the sensitivity of the results in Fig. \ref{fig:2} to these parameter choices. In the collisional growth mode (Fig. \ref{fig:s11}A), larger protoplanet mass increases the growth timescale, which means slower growth. In contrast, in the pebble accretion mode, larger protoplanet mass decreases the growth timescale, which means faster growth. In the earliest disk stages, however, only for protoplanet masses substantially higher than the mass of Ceres ($M_\mathrm{Ceres}$, $1 \, M_\mathrm{Ceres} \approx 1.6 \times 10^{-4} \, M_\mathrm{Earth}$) pebble accretion is faster than collisional accretion. Such high initial planetesimal masses are not produced in fluid dynamical simulations of the streaming instability mechanism \citep{2015SciA....115109J,2016ApJ...822...55S,2017ApJ...847L..12S,2019ApJ...883..192A,2019ApJ...885...69L,2021MNRAS.500..520R}, nor supported by asteroid belt size-frequency distribution reconstructions \citep{2009Icar..204..558M,2017Sci...357.1026D,2018Icar..304...14T}, or the Kuiper-belt impact cratering record \citep{2019Sci...363..955S}. In Reservoir I at around 1 Myr after CAI formation, if sufficiently massive embryos are present, the mass doubling timescale for pebble accretion onto Moon-like embryos is less than 1 Myr, which is the approximate time interval of a high pebble flux in the inner Solar System (cf. Fig. \ref{fig:2}). 

These results align with previous modeling results that suggest a heterogeneous transition of the dominant growth mode from collisional to pebble accretion \citep{2015SciA....115109J,2016A&A...586A..66V,2019A&A...624A.114L}. In Reservoir II the growth timescales for pebble accretion display a larger deviation with embryo mass relative to Reservoir I, however, similarly initial growth must proceed via planetesimal collisions to reach masses susceptible for pebble accretion to commence. The increase in accretion timescale for Reservoir I starting at $\approx$ 3 Myr after CAI formation suggests that pebble accretion becomes inefficient even for higher masses at late disk stages. Previous simulations with a truncated pebble flux in the inner Solar System \citep{2015PNAS..11214180L} and modeling work investigating super-Earth formation \citep{2019A&A...627A..83L} similarly suggest a high sensitivity of protoplanet growth by pebble accretion on the local pebble flux.

\subsubsection*{Open source software packages}

The numerical simulations and calculations in this work made use of \textsc{scipy} \citep{scipy2020}, \textsc{numpy} \citep{numpy2020}, \textsc{pandas} \citep{pandas:2010}, and \textsc{astropy} \citep{2018AJ....156..123A}.  Figures were produced using \textsc{matplotlib} \citep{Hunter:2007} and \textsc{seaborn} \citep{seaborn:2020}.

\subsection*{Supplementary Text}
\label{sec:constraints}

\subsubsection*{Evidence for compositional dichotomy \& accretion chronology}

There are compositional and mass budget trends with heliocentric distance \citep{1981PThPS..70...35H,1982Sci...216.1405G} for which various explanations have been proposed \citep{1993Sci...259..653G,2015Icar..258..418M,2016Icar..267..368M}. Mass budget calculations show $\approx 2 M_{\mathrm{Earth}}$ in solids in the inner and $\approx 445 \, M_{\mathrm{Earth}} = 1.4 \, M_{\mathrm{Jupiter}}$ in solids and gas in the outer Solar System planets \citep{2015Icar..258..418M,2018SSRv..214..101N}. Water abundance increases with orbital distance \citep{2015aste.book...13D,Raymond2020PlanetaryAstrobiology}. Hf-W dating of core segregation events \citep{2014Sci...344.1150K} and Pb-Pb ages of chondrules \citep{2017SciA....3E0407B} indicate that the planet-forming embryos of Earth and Mars started their accretion early during the disk phase at $\lesssim$1--2 Myr after CAI formation, but did not gain their full mass until $\gtrsim$ 30 Myr for Earth and $\approx$ 4--15 Myr for Mars \citep{2011Natur.473..489D,2010NatGe...3..439R,Marchi20SciAdv}, depending on the assumptions used in the Hf-W chronometry \citep{2007Icar..191..497N,2013SSRv..174...27M}. 

A hypothesis for the origin of the observed CC/NC dichotomy is an early formation of Jupiter \citep{2017PNAS..114.6712K,2018ApJS..238...11D,2018ApJ...854..164S}, but the Hf-W constraints on the accretion process substantially narrow the possible parameter space for a simultaneous rapid accretion of proto-Jupiter \citep{Alibert18NatAstron} and protracted growth of the inner terrestrial planets \citep{2018SSRv..214..101N,2020NatAs.tmp....8B}. An alternative physical mechanism for the separation of the inner and outer Solar System reservoirs is required if proto-Jupiter formed far from the inner Solar System and late during the disk stage. This is consistent with Jupiter's atmospheric volatile budget \citep{2019AJ....158..194O,2019A&A...632L..11B} and the asymmetric distribution of its trojan asteroids \citep{2019A&A...623A.169P}. The proposed filtering mechanism in the Jupiter barrier scenario is sensitive to the dynamical configuration of the gas giants \citep{2019AJ....158...55H} and the dust destruction efficiency in the outer pressure bump \citep{2019ApJ...885...91D}. It is also unclear by what mechanism gas accretion onto giant planets halts and if Jupiter would stop to grow at sufficient mass during the disk lifetime \citep{2019A&A...630A..82L} if Jupiter's core would have been formed $\lesssim$ 1 Myr after CAIs, as required for an early-formed Jupiter to explain the CC/NC dichotomy \citep{2017PNAS..114.6712K}. 

The Hf-W time constraints on embryo growth indicate early core-segregation events on the parent bodies of iron meteorites in the inner Solar System reservoir from $\approx$0--2 Myr after CAI formation \citep{2014Sci...344.1150K,2017PNAS..114.6712K,Hunt18EPSL,2019GeCoA.244..416P}, which implies the formation of these bodies sometime earlier, contemporaneous with or shortly after the formation of CAIs \citep{2012Sci...338..651C,2010NatGe...3..637B}. In contrast, core segregation in the outer Solar System reservoir (as witnessed by carbonaceous chondrites or differentiated material related to it) is delayed relative to the inner Solar System by $\approx$1.3 Myr (Tab. \ref{tab:ages}). Astronomical observations of protoplanetary disks reveal a variety of substructure \citep{2018ApJ...869L..41A}, and indicate early grain growth during the disk infall stage \citep{2018NatAs...2..646H,2020Natur.586..228S} and a decline of dust mass with time \citep{2010MNRAS.407.1981G,2016ApJ...831..125P,2018A&A...618L...3M,2020A&A...640A..19T}. Some disks have been observed with ringed substructure as close in as $\approx$1~au \citep{2016ApJ...820L..40A} or during the Class I disk stage \citep{2020Natur.586..228S}. However, most observed disks are several Myr old, and the orbital locations of gaps (observed at mm wavelength) range from tens to hundreds of au. These disk observations complement the geochemical chronology and suggest the onset of planetary accretion potentially as early as the Class 0 or Class I disk stage of the Solar System \citep{2010pdac.book..263P}.

\begin{table}[tb!]
\centering
\caption{\textsf{\textbf{Times of metal-silicate segregation and aqueous alteration in meteorites.} Times relative to CAI formation \citep{2012Sci...338..651C}. CM, CR, and CI values are weighted means from several meteorites \citep{2015NatCo...6.7444D,2017GeCoA.201..224J}. Used to generate Figs. \ref{fig:4}A and \ref{fig:5}A. References: (1) \citet{2017PNAS..114.6712K}, (2) \citet{Hunt18EPSL}, (3) \citet{2015NatCo...6.7444D}, (4) \citet{2013E&PSL.362..130F}, (5) \citet{2012NatCo...3..627F}, (6) \citet{2017GeCoA.201..224J}.}}
\begin{tabular}{lllll}
Meteorite group & Reservoir & $\Delta t_{\mathrm{CAI}} \pm 2 \sigma$ (Myr) & Ref. \\ \hline \hline
&&& \\
\emph{Metal-silicate segregation} &  &  &  \\ 

IC          & NC & 0.3 $\pm$ 0.5 & (1) \\
IIAB        & NC & 0.8 $\pm$ 0.5 & (1) \\
IIIAB       & NC & 1.2 $\pm$ 0.5 & (1) \\
IVA         & NC & 1.5 $\pm$ 0.6 & (1) \\
IIIE        & NC & 1.8 $\pm$ 0.7 & (1) \\
IAB         & NC & 6.0 $\pm$ 0.8 & (2) \\ \hline

IIIF        & CC & 2.2 $\pm$ 1.1 & (1) \\
IID         & CC & 2.3 $\pm$ 0.6 & (1) \\
IIF         & CC & 2.5 $\pm$ 0.7 & (1) \\
IIC         & CC & 2.6 $\pm$ 1.3 & (1) \\
IVB         & CC & 2.8 $\pm$ 0.7 & (1) \\
&&& \\

\emph{Aqueous alteration} &  &  & \\ 

OC          & NC & 2.4 $^{+1.8}_{-1.3}$  & (3) \\ \hline

Tagish Lake & CC & 4.1 $^{+1.3}_{-1.1}$  & (4) \\
CM          & CC & 4.1 $^{+0.5}_{-0.4}$  & (5) \\
CV          & CC & 4.2 $^{+0.8}_{-0.7}$  & (3) \\
CR          & CC & 4.2 $^{+1.8}_{-1.3}$ & (6) \\
CI          & CC & 4.4 $^{+0.6}_{-0.5}$ & (4) \\
CO          & CC & 5.1 $^{+0.5}_{-0.4}$  & (3) \\ 

\end{tabular}
\label{tab:ages}
\end{table}

The formation of secondary minerals and the high oxidation state of iron phases in CC parent bodies reveal that they initially formed with larger water budgets and subsequently dehydrated \citep{1989Icar...82..244G,2018SSRv..214...36A,2015aste.book..635K,2017GeCoA.211..115S}. Inner Solar System water requires several sources \citep{2014prpl.conf..835V,2019E&PSL.52615771M,PeslierISSI2018,2018SSRv..214...36A}. Similarities in noble gas abundances between carbonaceous chondrites and differentiated planetary bodies, and their nitrogen and hydrogen isotope ratios suggest CCs are the principal source of Earth's volatile elements \citep{2012Sci...337..721A,2012E&PSL.313...56M}, in addition to some potential nebular ingassing \citep{2018JGRE..123.2691W,2019Natur.565...78W}. However, uncertainties persist in the timing of arrival of the volatile source bodies and their nature and prior internal evolution \citep{2014Sci...346..623S,2019SciA....5.3669G,2019GeCoA.260..204S,Piani+2020}. Deuterium abundances indicate an early contribution from interstellar volatile ices \citep{2014Sci...345.1590C,2018NatAs...2..317P} or extensive iron oxidation \citep{2017GeCoA.211..115S}. Estimates of the initial water abundance in meteoritic parent bodies range from several per cent \citep{2018E&PSL.482...23M} to up to 50 per cent water by mass \citep{2003ApJ...591.1220L,2014prpl.conf..835V}. Traces of fluid flow are widespread in the meteorite record, even in NC bodies \citep{2016M&PS...51.1886L,2018GeCoA.240..293L}. The carbonaceous chondrites exhibit the widest range of water abundances \citep{2012Sci...337..721A,2018SSRv..214...36A}, but some NC chondrites also display substantial abundances with up to 6000 ppm H$_\mathrm{2}$O \citep{2019E&PSL.52615771M}. The range in measurements indicates that the delivery of water and other highly volatile elements to the inner terrestrial planets was subject to the stochastic and evolving nature of the accretion process.

\subsubsection*{Nucleosynthetic isotope variability in planetary materials}
\label{sec:dichotomy}
\begin{figure*}[tbh]
   \centering
   \includegraphics[width=0.95\textwidth]{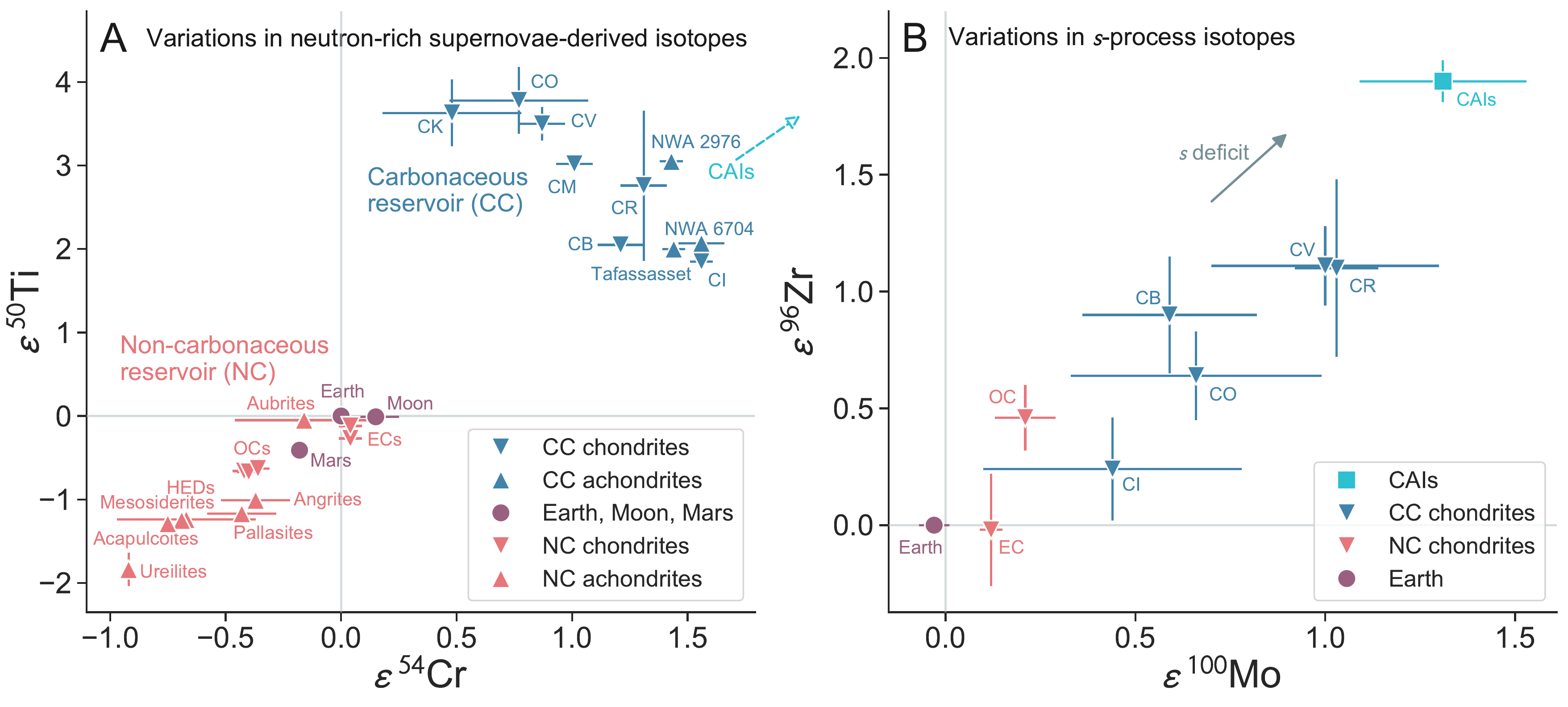}
   \caption{\textsf{\textbf{Nucleosynthetic isotope anomalies in Solar System planetary materials.} (\textbf{A}) Nucleosynthetic Ti and Cr isotope data in $\varepsilon$ notation (parts per 10,000 deviation of the ratio from the respective terrestrial standards). Average values are given for various chondrite and achondrite groups (symbols). Tafassasset, NWA 2976 and NWA 6704 refer to individual meteorites. Earth, Moon and Mars reflect data from Earth's mantle, lunar Apollo samples, and Martian meteorites for bulk compositions, respectively. HEDs (Howardite-Eucrite-Diogenites) are meteorites considered to originate from Vesta. CC = carbonaceous chondrites and related groups (blue), NC = non-carbonaceous groups (red), ECs = enstatite chondrites, OCs = ordinary chondrites. Data from \citep{2007ApJ...655.1179T,2008E&PSL.266..233L,2009ApJ...702.1118L,2011ApJ...735L..37L,2019GeCoA.245..577S,2015GeCoA.156....1G}. 2-$\sigma$ uncertainties are indicated with red and blue lines on the symbols. The dashed turquoise arrow points toward the location of CAIs in Ti-Cr space, off the displayed axes limits. Light gray horizontal and vertical lines center on the axes origin. (\textbf{B}) Nucleosynthetic Zr and Mo isotope data of various chondrite classes. Same notation and colors as in panel \textbf{A}. Earth is most enriched in $s$-process isotopes compared to all other analyzed materials. Zr data from \citep{2013ApJ...777..169A,2015GeCoA.165..484A}, Mo data from \citep{2011E&PSL.312..390B,2018ApJ...862...26R}. Both panels adapted from \citet{Mezger+2020}.}}
   \label{fig:s12}
\end{figure*}
Distinct nucleosynthetic compositions have been identified for each meteorite parent body and planet, for which samples are available for analyses (i.e., Earth and Mars). These nucleosynthetic variations are caused by the heterogeneous distribution of presolar dust, carrying a characteristic isotopic composition that reflects its stellar formation site \citep{2018PrPNP.102....1L}. This stardust contributed $\approx$3\% of the total dust in the protoplanetary disk \citep{2017NatAs...1..617H}. Stardust was incorporated to different extents into each planetary body and led to their distinct isotope compositions \citep{2019NatAs.tmp....2E}. Two first order types of anomalies can be distinguished (Fig. \ref{fig:s12}): variations in \emph{s}-process isotopes \citep{2019NatAs.tmp....2E,2018ApJ...862...26R,2015GeCoA.165..484A,2011E&PSL.312..390B} and variations in neutron-rich isotopes such as $^{48}$Ca, $^{50}$Ti and $^{54}$Cr \citep{2007ApJ...655.1179T,2008E&PSL.266..233L,2009Sci...324..374T}. The latter indicate nucleosynthesis in a supernova environment. The variability in neutron-rich isotopes shows a distinct difference between carbonaceous chondrites (CC) and the remaining Solar System material available for laboratory studies (i.e. non-carbonaceous chondrites, Mars and Earth, NC).

The origin of this isotopic dichotomy has broadly been attributed to two processes: \emph{(a)} a change in composition of the material infalling onto the disk because of a heterogeneous distribution of stardust in the molecular cloud \citep{2019E&PSL.511...44N,2019ApJ...884...32J} or \emph{(b)} thermal processing and selective destruction of dust during the infall or in the disk itself. The latter process selectively removed specific carrier phases of nucleosynthetic anomalies \citep{2009Sci...324..374T,2016PNAS..113.2011V} and assumes a homogeneous and well-mixed molecular cloud. In contrast to the neutron-rich nucleosynthetic variability (e.g., in $^{50}$Ti or $^{54}$Cr), the \emph{s}-process variations show a single array with no clear dichotomy gap (Fig. \ref{fig:s12}B) \citep{2015GeCoA.165..484A,2019NatAs.tmp....2E}, with the exception of Mo isotopes \citep{2017E&PSL.473..215P,2017PNAS..114.6712K}. Based on the \emph{s}-process variations, dust grown in the interstellar medium (ISM) may have been destroyed by thermal processing leading to the \emph{s}-process variations \citep{2019NatAs.tmp....2E}. This ISM-dust contributes $\approx$97\% of the dust in the protoplanetary disk \citep{2017NatAs...1..617H} and has an average Solar System composition.

In our proposed scenario of early Solar System accretion (Fig. \ref{fig:6}), a change in the composition of the infalling material (model \emph{(a)}) can explain the initial establishment of the nucleosynthetic dichotomy \citep{2019E&PSL.511...44N,2019ApJ...884...32J}. Planetary objects in the Reservoir I region are dominated by material that is mainly fed to the inner disk during later infall stages and thus influences the composition of these planetesimals, while Reservoir II preferentially samples older material from before the infall changed composition. This older material, which includes CAIs, is transported to the outer part of the disk during the Class I disk phase through viscous spreading and diffusional transport \citep{2018ApJ...867L..23P,2012M&PS...47...99Y,2019E&PSL.511...44N,2019ApJ...884...32J}. However, in contrast to previous suggestions \citep{2017PNAS..114.6712K,2018ApJS..238...11D,2018ApJ...854..164S}, our proposed scenario does not require the presence of proto-Jupiter to block the addition of outer Solar System materials to Reservoir I in the inner Solar System. Instead, both the chemical and isotopic dichotomy are an inherent consequence of the two subsequent episodes of planetesimal formation and the pebble flux suppression due to the dust pile-up along the drifting snow line.

Similarly, thermal processing of infalling dust (model \emph{(b)}) is consistent with our proposed model and could have generated or contributed to the observed nucleosynthetic heterogeneities. However, selective destruction of dust in the disk itself is less likely, because in our model most of the thermal processing of inner Solar System material occurs inside of the planetesimals. This processing should not affect the nucleosynthetic anomalies because these are mainly carried by refractory elements \citep{2020GeCoA.274..286T}.

\vfill \pagebreak

\bibliography{references.bib}
\bibliographystyle{aasjournal}

\end{document}